\documentclass[11pt,aps,preprintnumbers,nofootinbib,superscriptaddress,notitlepage,longbibliography]{revtex4-2}

\usepackage[T1]{fontenc}
\usepackage{slashed}
\usepackage[english]{babel}
\usepackage{epsfig}
\usepackage{amsmath,amsfonts,physics,tensor,dsfont,amssymb}
\usepackage{graphicx}
\usepackage{verbatim}
\usepackage{csquotes}
\usepackage{simplewick}
\usepackage{appendix}
\usepackage[dvipsnames,svgnames]{xcolor}
\usepackage[normalem]{ulem}
\usepackage{hyperref}
\usepackage{cleveref}
\usepackage{orcidlink}
\usepackage{bbm}
\usepackage{soul}

\usepackage[framemethod=default]{mdframed}
\newmdenv[skipabove=7pt,
skipbelow=7pt,
rightline=false,
leftline=false,
topline=false,
bottomline=false,
backgroundcolor=gray!10,
linecolor=gray,
innerleftmargin=5pt,
innerrightmargin=5pt,
innertopmargin=5pt,
innerbottommargin=5pt,
leftmargin=0cm,
rightmargin=0cm,
linewidth=4pt]{eBox}

\newcommand{\covDLR}{\overset{\leftrightarrow}{D}}

\newcommand{\dfx}{\mathrm{d}^4x}

\newcommand{\md}{\mathrm{d}}

\definecolor{IAN}{RGB}{1,80,158}

\begin{document}

\title{Apparent Lorentz violation from disformally coupled ultralight dark matter
}

\author{\textsc{Guillem Dom\`enech\,\orcidlink{0000-0003-2788-884X}}}
    \email{{guillem.domenech}@{itp.uni-hannover.de}}
\affiliation{Institute for Theoretical Physics, Leibniz University Hannover, Appelstraße 2, 30167 Hannover, Germany.}
\affiliation{ Max Planck Institute for Gravitational Physics, Albert Einstein Institute, 30167 Hannover, Germany.}

\author{\textsc{Alexander Ganz\,\orcidlink{0000-0001-7939-9058}}}
    \email{{alexander.ganz@itp.uni-hannover.de}}
    \affiliation{Institute for Theoretical Physics, Leibniz University Hannover, Appelstraße 2, 30167 Hannover, Germany.}

\author{\textsc{Fiona Kirk\,\orcidlink{0000-0002-2234-5216}}}

\email{{fiona.kirk}@{itp.uni-hannover.de}}
    \affiliation{Physikalisch-Technische Bundesanstalt, Bundesallee 100, Braunschweig, 38116, Germany.}

    \affiliation{Institute for Theoretical Physics, Leibniz University Hannover, Appelstraße 2, 30167 Hannover, Germany.}
\affiliation{Department of Particle Physics and Astrophysics, Weizmann Institute of Science, Rehovot, Israel 7610001}
    
\author{\textsc{Nathaniel Sherrill\,\orcidlink{...}}}
\email{{nathaniel.sherrill}@{itp.uni-hannover.de}}
    \affiliation{Institute for Theoretical Physics, Leibniz University Hannover, Appelstraße 2, 30167 Hannover, Germany.}

\author{\textsc{Apostolos Tsabodimos\,\orcidlink{0009-0001-0230-5647}}}
    \email{{apostolos.tsabodimos}@{stud.uni-hannover.de}}
    \affiliation{Institute for Theoretical Physics, Leibniz University Hannover, Appelstraße 2, 30167 Hannover, Germany.}
    \affiliation{Center for Physical Sciences and Technology, Sauletekio 3, 10257 Vilnius, Lithuania}

    \begin{abstract}
  
We study
the impact of general disformal metric transformations on fermions, which shift the gravitational metric by an additional rank-2 tensor. This tensor can in principle be constructed from scalar-field gradients, vector fields, or field-strength contractions. 
We show this transformation results in the conventional Dirac action being modified by additional kinetic and axial-current  couplings that are quadratic in the shifted field. When the field sourcing the metric shift takes on a non-trivial background value, apparent Lorentz-violating effects can result, which we identify as terms in an effective field theory. 
Assuming the well-motivated cases of scalar and vector ultralight dark matter, 
we demonstrate that experimental tests of rotation and boost violation imply constraints on the additional kinetic coupling. Even under conservative assumptions, the constraints for vector ultralight dark matter are extremely stringent.
    \end{abstract}

\maketitle

    \section{Introduction  \label{sec:Intro}}

An increasingly popular approach to search for physics beyond the Standard Model (SM) and General Relativity involves introducing a weakly coupled bosonic field to ordinary matter. When the field is much lighter than matter and contains an orientation, either via a spacetime derivative or other spacetime indices, \textit{apparent Lorentz violation} can result~\cite{Kostelecky:2002ca,Nelson:2011sf}. 

Leveraging the sensitivity of Lorentz-invariance tests~\cite{auth:KosteleckTables2025}, several scenarios of this type have been probed, including axion and scalar ultralight dark matter (ULDM)~\cite{Hardy:2024fen,Cordero:2024nho,Cordero:2024hjr,Jiang:2024agx,Zhang:2025zcq,Graham:2017ivz}, electromagnetic axion-like interactions~\cite{Carvalho:2022mgk,Borges:2013eda,Karki:2020rgi} and decays~\cite{Gupta:2022qoq}, neutrino-dark energy~\cite{Klop:2017dim} and dark matter~\cite{Lambiase:2023hpq}, and ultralight boson-fermion couplings~\cite{Carenza:2025jwn}. In the present work, we contribute to this relatively new and growing body of literature by showing that tests of Lorentz invariance strictly constrain a class of so-called \textit{disformal fermion couplings}. 

Modern experimental searches for Lorentz violation often adopt the effective field theory (EFT) framework known as the Standard-Model Extension (SME)~\cite{Colladay:1996iz,Colladay:1998fq,Kostelecky:2003fs}.
The SME action, $S_{\rm SME} = S_{\rm SM} + S_{\rm EH} + S_{\rm LV}$, contains the usual SM and Einstein-Hilbert actions, plus all possible Lorentz-violating terms consistent with general coordinate invariance. 
A given term in $S_{\rm LV}$ involves a coordinate-independent contraction of an EFT controlling coefficient, or \textit{SME coefficient}, with the conventional SM and gravitational fields and their covariant derivatives.  
Numerous constraints have been placed on the SME coefficients, as tabulated annually in the \textit{Data Tables for Lorentz and CPT violation}~\cite{auth:KosteleckTables2025}. A subset of them, typically ones with odd numbers of indices, also control CPT violation. 
Reviews of the SME include, e.g., Refs.~\cite{Bluhm:2005uj,Tasson:2014dfa}.

Two theoretical scenarios for Lorentz violation are often considered: \textit{explicit} or \textit{spontaneous} violation~\cite{Kostelecky:2003fs}. With explicit violation, the symmetry is violated nondynamically via an unspecified mechanism. In contrast, with spontaneous violation, the symmetry
is dynamically broken 
when a tensor-valued field acquires a nonzero vacuum expectation value (\emph{vev}). 
Examples of the latter scenario include string models~\cite{Kostelecky:1988zi} and higher-derivative gravity theories~\cite{Brax:2012hm,Sawicki:2024ryt}, such as of the Horndeski~\cite{Horndeski:1974wa,Gao:2014fra,Gao:2014soa,Gleyzes:2014dya,Langlois:2015cwa,DeFelice:2018ewo,Kobayashi:2019hrl} and Proca~\cite{Kostelecky:2003fs,Heisenberg:2014rta,Heisenberg:2016eld,Kimura:2016rzw,deRham:2020yet,Delhom:2022xfo} type, which can emerge due to dimensional reduction~\cite{Trodden:2011xh,Koivisto:2013fta,Dvali:2000hr,Mavromatos:2022xdo}.
Though spontaneous violation is arguably more natural, we emphasize that models of explicit and spontaneous violation are, by construction, encompassed within the model-independent SME. A comprehensive comparison of the two scenarios can be found in Ref.~\cite{Bluhm:2023kph}.

An attractive way of introducing spontaneous Lorentz violation in higher-derivative theories is through Bekenstein's proposal of \textit{disformal transformations}---a causal-preserving transformation which modifies the light cone~\cite{auth:BekensteinOriginalDisformal}. 
Extending the idea of a conformal transformation, 
the disformal transformation
acts like an anisotropic lens---stretching and compressing the metric in specific directions---and leads to exotic phenomena such as vacuum Čerenkov radiation and vacuum bremsstrahlung~\cite{vandeBruck:2016cnh}.

After Bekenstein’s seminal work, it was demonstrated that scalar–tensor theories remain closed under disformal transformations~\cite{Bettoni:2013diz,Zumalacarregui:2013pma,Ezquiaga:2017ner,Crisostomi:2016czh,BenAchour:2016cay,BenAchour:2016fzp,deRham:2016wji}, shifting interactions between the gravity and matter sectors via disformal couplings~\cite{Brax:2012hm}.
Concomitantly, disformal couplings of dark matter (DM) and dark energy (DE) have been investigated in: astrophysics~\cite {Sakstein:2014isa,Brax:2018bow}, cosmology~\cite{Zumalacarregui:2010wj,vandeBruck:2015ida,Brax:2020gqg, Koivisto:2008ak,Kaloper:2003yf}, black-hole physics~\cite{Minamitsuji:2020jvf,Babichev:2024hjf} and particle physics~\cite {Brax:2014vva,Brax:2015hma}).
However, to our knowledge, this literature has only limited attention to leading-order interactions of a single scalar field. 

In this work, we extend previous literature in three ways: first, we generalize the disformal transformation in the fermion sector with an arbitrary two rank tensor; second, we apply our results by assuming scalar and vector ULDM couplings to the SM, showing that apparent Lorentz-violating effects are generated; third, we utilize experimental tests of Lorentz violation to place constraints on the disformal couplings. Though our results for scalar ULDM  are not competitive, we find those for vector ULDM, benefitting from a reduced operator dimensionality in the action, are significantly enhanced.

The next sections are organized as follows. Given the variety of topics discussed, we first opt in Sec.~\ref{sec:preliminaries} to provide a light introduction to disformal theory and its connection to Lorentz violation in the SME framework. In Sec.~\ref{sec:MainBody}, we generalize disformal transformations in the fermion sector and in Sec.~\ref{sec:Examples} highlight a few examples, connecting to prior work where applicable. Next, in Sec.~\ref{sec:applications}, we apply these examples to scalar and vector ULDM and tests of rotation-invariance violation. Finally, in Sec.~\ref{sec:conclusions} we provide discussions and conclusions. Note that many calculational details, in particular for vector fields and tests of boost-invariance violation, are provided in the appendices. We work in reduced Planck units throughout, that is $\hbar=c=8\pi G=1$, and in the $+---$ signature of the metric.

    \section{Disformal theory and Lorentz violation \label{sec:preliminaries}}
   
    We start with a short overview, motivation, and specification of the framework under consideration. We hope this will be useful to the reader before we delve into details, which range from purely mathematical aspects to particle phenomenology and experimental constraints. 

\subsection{Disformal theory}
    Our starting point is the minimal assumption that modifications of gravity only occur through an effective metric $\tilde{g}$ to which matter fields couple minimally.\footnote{This means that the matter fields do not couple directly to any new fields. Consequently, all matter fields couple in the same way to the gravitational field, i.e., the weak equivalence principle is satisfied.} 
    This is in line with Bekenstein's original idea \cite{auth:BekensteinOriginalDisformal}, where (in Bekenstein's notation) one has a ``gravitational'' metric $g$ and a ``physical'' one $\tilde{g}$. The dynamics of the ``gravitational'' metric are described by the Einstein-Hilbert action plus an additional canonical field whose energy density dominates over the other components. Matter is minimally coupled to the ``physical'' metric $\tilde{g}$, which we henceforth call the matter metric. 
    An analogous construction can be considered in the photon sector if we introduce a photon metric $\breve{g}$.
    The total action then reads
    \begin{align}\label{eq:firstaction}
    S=&\int \dfx \sqrt{-g}\Big\{\frac{1}{16\pi G}R +{\cal L}(\phi,B_\mu)\Big\}\nonumber\\&
    +\int \dfx \sqrt{-\tilde g}\Big\{\frac{i}{2} (\overline{\tilde \Psi} \gamma^\mu \tilde{\overset{\leftrightarrow}{D}}_\mu \tilde \Psi) - \overline{\tilde \Psi} m \tilde \Psi\Big\} - \frac{1}{4} \int \dfx \sqrt{-\breve g} \, \breve g^{\mu\nu} \breve g^{\alpha\beta} F_{\mu\alpha} F_{\nu\beta}\,.
    \end{align} 
    The first line includes the 
    Ricci curvature $R$ and the Lagrange density ${\cal L}(\phi,B_\mu)$ of a dark scalar field $\phi$ or a dark massive vector field $B_\mu$ (not necessarily a gauge field).\footnote{Identification as DM is not strictly necessary, but is made given it is our target application.} In the second line, $\tilde\Psi$ denotes a fermion field associated to the metric $\tilde{g}$ and coupled to the photon field $A_\mu$ through ${{D}}_\mu={{\nabla}}_\mu+ieA_\mu$, where ${\nabla}_\mu$ is the covariant derivative and 
    $\overline{ \Psi} \gamma^\mu {\overset{\leftrightarrow}{D}}_\mu  \Psi \equiv \overline{ \Psi} \gamma^\mu {{D}}_\mu  \Psi-({D}^*_\mu\overline{ \Psi}) \gamma^\mu  \Psi$, and $ F_{\mu\nu}=\partial_{\mu} A_{\nu}-\partial_{\nu} A_{\mu}$. The properties of ${\nabla}_\mu$ are given in Sec.~\ref{sec:MainBody}.

    Although the action \eqref{eq:firstaction} might at first appear confusing, ultimately one can write it in terms of a single effective metric. We do so shortly in Eq.~\eqref{eq:secondaction}. But, first, we assume the effective metric $\tilde g$ is related to the gravitational metric $g$ via a \textit{disformal transformation}\footnote{With \emph{transformation} we mean a given relation between two metrics, here between the gravitational metric $g_{\mu\nu}$ and the matter metric $\tilde{g}_{\mu\nu}$.} given by    \begin{align}\label{eq:disformalpreliminary}
    \tilde g_{\mu\nu}=&C^2\left(g_{\mu\nu}+{\cal D} B_\mu B_\nu\right)=C^2(\tensor{\delta}{^\alpha_\mu}+D B^{\alpha} B_\mu)(\tensor{\delta}{^\beta_\nu}+D B^{\beta} B_\nu)g_{\alpha\beta}\notag\\
    =& C^2 \left(g_{\mu\nu} + D\left(2 + DX\right) B_\mu B_\nu\right)\,,
    \end{align}
    where $X=B_\mu B^\mu$, and $C$ and ${\cal D}=D(2+DX)$ are the so-called \emph{conformal} and \emph{disformal couplings}. 
    In the second step of \eqref{eq:disformalpreliminary}, we wrote the disformal transformation in a form that helps to connect with the transformation of the local tetrads (also known as vierbeins). To preserve the metric signature, one requires $1+{\cal D}X>0$ (or, equivalently, $|1+DX|>0$). To maintain a positive definite volume element, one further requires $1+DX>0$ \cite{Ezquiaga:2017ner}. Note that although we focus here on vector disformal transformations, we treat the general case shortly in Sec.~\ref{sec:MainBody}.

    Since we will be focusing on the fermion sector, and  $D$ is the relevant disformal coupling for fermions, we refer to $D$ as the disformal coupling in the following. 
    Both $C$ and $D$ can, a priori, be arbitrary scalar functions of $X$.
    Note for this particular choice of $X$ that $C$ has mass dimension zero, while $\mathcal{D}$ and $D$ have a mass dimension $-2$.
    For the case of a scalar field, one simply replaces $B_\mu\to \nabla_\mu\phi$. Then $C$, ${\cal D}$ and $D$ have mass dimensions zero, $-4$ and $-4$, respectively, and may depend on $\phi$ and its first derivatives. 
     A similar expression to Eq.~\eqref{eq:disformalpreliminary} holds for the photon metric $\breve g_{\mu\nu}$ with different $C$, ${\cal D}$ and $D$ functions. We discuss more general couplings, as, e.g., derivatively coupled dark vector fields 
     in a separate publication \cite{auth:us2}.

     Before proceeding, we fix the action and write it explicitly in terms of a single metric.
     We make use of Gravitational Wave (GW) observations from a neutron star merger with an electromagnetic counterpart \cite{LIGOScientific:2017vwq}, which have severely constrained the velocity of GWs relative to photons (see also Ref.~\cite{Brax:2013nsa} for other cosmological tests), to set $\breve g_{\mu\nu}=g_{\mu\nu}$, but leave an arbitrary coupling to fermions. 
     We find that the 
     action after the disformal transformation reads
     \begin{align}
     \label{eq:secondaction}
    S=&\int \dfx 
    \sqrt{-g}\Big\{\frac{1}{16\pi G}R +{\cal L}(\phi,B_\mu)\Big\}- \frac{1}{4} \int \dfx  \sqrt{-g} \,  g^{\mu\nu}  g^{\alpha\beta} F_{\mu\alpha} F_{\nu\beta}\nonumber\\&+\int \dfx
    \sqrt{-g}\Big\{ \frac{i}{2} \left(\tensor{\delta}{^\mu_\nu}-\frac{D B^\mu B_\nu}{1+D X}\right)\overline \Psi \gamma^\nu \overset{\leftrightarrow}{D}_\mu \Psi - \overline{ \Psi} m C \Psi- \frac{1}{4}\frac{ D^2 X }{1+ D X}  B^{\mu}  {\cal F}^{\star}_{\mu\nu}\overline \Psi  \gamma^5\gamma^\nu  \Psi\Big\}\,,
    \end{align} 
    where $\mathcal{F}_{\mu\nu}^\star$ is the (dual) field-strength tensor of the dark vector $B_\mu$:
    \begin{align}
    {\cal F}^\star_{\mu\nu}=\frac{1}{2}\epsilon_{\mu\nu\alpha\beta}{\cal F}^{\alpha\beta}\quad{\rm with}\quad {\cal F}_{\alpha\beta} = \partial_\alpha B_\beta - \partial_\beta B_\alpha\,.
    \end{align}
    Note the axial coupling in the last term of Eq.~\eqref{eq:secondaction}, which may be understood as originating from a spin-vorticity coupling \cite{PhysRevB.96.020401}. 
    
    If $B_\mu$ represents a non-trivial background field, the action in Eq.~\eqref{eq:secondaction} can induce violations of Lorentz invariance in the fermion sector. 
    A common example in the literature is that of a scalar field $\phi$ whose gradient couples as $B_\mu $ in Eq.~\eqref{eq:secondaction} and varies on cosmological scales, i.e. $B_\mu \to \nabla_\mu\phi=\partial_0\phi \,\delta_\mu^0$.
    Moreover, if we treat $B_\mu$ as a background, the action in Eq.~\eqref{eq:secondaction} can be identified as a subset of the SME, as we show in the following section.

 \subsection{Lorentz violation}

The SME terms of interest are contained in the so-called minimal quantum-electrodynamics (QED) sector coupled to gravity~\cite{Kostelecky:2003fs,Kostelecky:2010ze}. The relevant fermion action is 
\begin{align}
\label{LSME}
S_{\rm SME} \supset 
\int \dfx\; \left( \tfrac{1}{2}iee^\mu_{\hphantom{\mu}A}\bar\Psi \Gamma^A \covDLR_\mu \Psi  -  e\bar\Psi M\Psi \right)\,,
\end{align}
where 
$e^\mu_{A}$ is the tetrad (vierbein) with determinant $e=\det(e_\mu^A)=\sqrt{-g}$.  
$\Gamma^A$ and $M$ are generalized Dirac and mass matrices, respectively, and contain
\begin{align}
\label{GammaM}
&\Gamma^A  \supset \gamma^A - c_{\mu\nu}e^{A\nu }e^\mu_{B\hphantom{\mu}}\gamma^B, \\
& M \supset m_\Psi 
+ b_\mu e^{\mu}_{A}\gamma_5\gamma^A\,,  \nonumber
\end{align}
where $m_\Psi$ is the fermion mass and $c_{\mu\nu}, b_\mu$ are SME coefficients parametrizing Lorentz violation.
The hermiticity of~\eqref{LSME} follows by assuming the coefficient components are real. The $c_{\mu\nu}$ coefficient is mass dimension zero and may be taken as symmetric and traceless, implying nine physical components~\cite{Kostelecky:2010ze}. 
Since the field operator associated with $b_\mu$ is odd under a CPT transformation, this coefficient of mass dimension +1 also controls CPT violation. 

By comparing Eq.~\eqref{LSME} to the disformal action~\eqref{eq:secondaction}, we establish the correspondence
\begin{equation}
        \tensor{c}{^\mu_\nu} \leftrightarrow  \frac{D B^\mu B_\nu}{1+DX}\,, \qquad
        b_\mu \leftrightarrow \frac{1}{4}\frac{D^2 X}{1+DX}B^\nu \mathcal{F}^\star_{\nu\mu}, \label{eq:correspondence}
\end{equation}
demonstrating that disformal fermion couplings are analogous to a subset of SME effects. The presence of a dynamical field $B_\mu$ sourcing the coefficients invites comparison to the scenario of spontaneous Lorentz violation, as investigated in Ref.~\cite{Delhom:2022xfo}. 
However, we stress that $c_{\mu\nu}$ and $b_\mu$ are defined as EFT parameters with prescribed unconventional properties under particle Lorentz transformations. They need not necessarily be described by underlying fields with (or without)  nonzero vevs, as the scenario of explicit violation briefly mentioned in  Sec.~\ref{sec:Intro} demonstrates. For these reasons,    we prefer to delay arguing a strict equality in~\eqref{eq:correspondence} until discussing our specific application in Sec.~\ref{sec:applications}.

    \section{General Disformal Coupling to Matter  \label{sec:MainBody}}

    Let us investigate the Dirac Lagrangian under the action of a general disformal transformation. In Sec.~\ref{sec:Examples}, we will focus on the scalar and vector disformal cases. Since we work with fermions in a curved background, we employ the tetrad formalism with differential forms, see, e.g., Refs.~\cite{auth:EzquiagaFieldRedefinitions,Gauthier:2009wc}.
    As most of the literature uses tensor component notation, we present the corresponding procedure in App.~\ref{sec:Transformation_Tensor_Components}. 
    
    Before going into the details, we lay out our notation throughout the rest of this work. We refer to lower-case Greek letters ($\mu$, $\nu$, $\rho$, etc.) as the spacetime indices, lowered and raised by the spacetime metric $g_{\mu\nu}$. Local indices are indicated by upper case Latin letters ($A$, $B$, $C$, etc.) with the corresponding metric $\tensor{\eta}{_A_B}$. The tetrad 1-forms are denoted as ${\theta^A} := \tensor{e}{^A_\mu} \md x^\mu$, whilst their duals as ${e_A} :=\tensor{e}{_A^\mu} \partial_\mu$, and by definition satisfy $\theta^A ( e_B) = \tensor{\delta}{^A_B}$.
    The action of the covariant derivative on a Dirac spinor is given by
    \begin{align}
        \nabla_\mu \Psi = \left( \partial_\mu + \frac{i}{2} \tensor{\omega}{_A_B_\mu} \Sigma^{AB} \right) \Psi\quad , \quad \nabla^*_\mu \overline\Psi = \left( \partial_\mu - \frac{i}{2} \tensor{\omega}{_A_B_\mu} \Sigma^{AB} \right) \overline\Psi\,,
    \end{align}
    where $\tensor{\omega}{_A_B_\mu}$ are the components of the Cartan-structure 1-form $\tensor{\varpi}{_A_B}$ (spin connection coefficients), namely
    \begin{equation}\label{eq:Connection_Coefficients_Definition}
           \tensor{\varpi}{_A_B}= \tensor{\omega}{_A_B_\mu} \md x^\mu = \tensor{\omega}{_A_B_C}\theta^C\,,
    \end{equation}
    and we use the antisymmetric Dirac element 
    \begin{align}
        \Sigma^{AB} = \frac{i}{4}[\gamma^A,\gamma^B]\,.
    \end{align}

    \subsection{Disformal transformation of the tetrad and spin connection}

    Introducing the tetrad frames $\{\tilde{\theta}^A\}$ and $\{\theta^A\}$, we define the general disformal transformation via
        \begin{align}\label{eq:ThetaDisformal}
            & \tilde{\theta}^A = C \left(\tensor{\delta}{^A_B}+ \tensor{L}{^A_B}\right)\theta^B \equiv C \tensor{\mathcal{U}}{^A_B}\theta^B\,,
            \\
            & \tilde e_A = \frac{1}{C} \left( \tensor{\delta}{^B_A}  +  \tensor{P}{^B_A} \right) e_B \equiv \frac{1}{C}\tensor{(\mathcal{U}^{-1})}{^B_A} e_B\,.
            \label{eq:ThetaDisformalInv}
        \end{align}
 The latter can be rendered symmetric and traceless without loss of generality, as we show in App.~\ref{app:Lorentz_transformation}. However, we find it more convenient to proceed as is. For comparison with the literature, the metric transformation corresponding to the tetrad change \eqref{eq:ThetaDisformal} reads
        \begin{align}
            &\tilde g =  \eta_{AB} \tensor{\tilde{\theta}}{^A}\otimes\tensor{\tilde{\theta}}{^B}  \nonumber= C^2 \left( \eta_{AB} + 2 L_{(AB)} + L_{AC} \tensor{L}{^C_B} \right) \theta^A\otimes  \theta^B\,.
            \label{eq:Metric_Disformal_transformation_General}
        \end{align}
    Lastly, the impact of the transformation on the corresponding volume form is computed using Eq.~\eqref{eq:ThetaDisformal}, namely 
        \begin{align}
           \sqrt{ -\tilde g} \dd[4]x = & \frac{1}{4!} \tensor{\epsilon}{_A_B_C_D} \tilde \theta^A\wedge \tilde \theta^B \wedge \tilde \theta^C \wedge \tilde \theta^D = \sqrt{-g} \,C^4\det \mathcal{U} \md^4x  \,.
       \end{align}
   
    Computing the transformation of the dual tetrad \eqref{eq:ThetaDisformalInv} corresponds to writing $P=[\tensor{P}{^A_B}]$ as a function of $L$. Assuming that the transformation is invertible, we find, using the Cayley-Hamilton theorem, that 
        \begin{equation}
            \label{eq:Expression_P}
                P =  \frac{1}{\det\mathcal{U}}\left[ -  L^3 + ( 1 + \Tr L)  L^2 + b_1 L + b_0 \, \mathds{1} \right] \,,
            \end{equation}
        where
        \begin{align}
            b_1 =&  - \frac{1}{2} \left( 2 + \Trace L (2 + \Trace L) - \Trace L^2 \right)  ~, \\
            b_0 =& \frac{1}{24} \left( - ( \Tr L)^4  - 8 \Tr L \Tr L^3 + 6 \Tr L^4 + 6 (\Tr L)^2 \Tr L^2 - 3  (\Tr L^2 )^2 \right) ~, 
        \end{align}
        and 
            \begin{align}
                  \det \mathcal{U} =& 1 + \Tr L  \left( 1 + \frac{1}{2} \Tr L + \frac{1}{6} ( \Tr L)^2 + \frac{1}{24} ( \Tr L)^3 \right) + \frac{1}{3} \Tr L^3  ( 1 + \Tr L) - \frac{1}{4} \Tr L^4  \nonumber \\
                & + \frac{1}{8} \Tr L^2 \left( -4 - 4 \Tr L - 2 (\Tr L)^2 + \Tr L^2 \right) \,.
              \end{align} 

        Having established the transformation properties of the tetrads, we turn to
        the spin connection coefficients. Following Ref.~\cite{auth:EzquiagaFieldRedefinitions}, we work with the Cartan structure 1-form postulating
            \begin{equation}\label{eq:CartanTranPost}
                \tensor{\tilde{\varpi}}{_A_B} = \tensor{\varpi}{_A_B} + \tensor{Z}{_A_B}\,,
            \end{equation}
        where $\tensor{Z}{_A_B}$ is a 1-form antisymmetric in $A$ and $B$. Using the torsionless condition, we find\footnote{Explicitly, we require that the torsion 2-form vanishes, that is
            \begin{equation}\label{eq:Torsion}\nonumber
                \tilde{T}^A = \tilde{\mathfrak{D}}\tilde{\theta}^A {=}0~,
            \end{equation}
        where $\mathfrak{D}$ is the exterior covariant derivative. 
        This translates into a condition for $\tensor{Z}{_A_B}$ given by
            \begin{equation}\label{eq:Z_equation}\nonumber
                \mathfrak{i}_C \tensor{Z}{_A_B} \tensor{\mathcal{U}}{_B_D} - \mathfrak{i}_D \tensor{Z}{_A_B} \tensor{\mathcal{U}}{_B_C}  =   \mathfrak{i}_D\mathfrak{i}_C S_A\,,
            \end{equation}
        where we used the shorthand notation $\mathfrak{i}_C = \mathfrak{i}_{e_C}$ for the interior derivative. We also define the two-form $S_A$ via
            \begin{equation}\label{eq:S_Local_Definition}\nonumber
                S_A = - \frac{1}{C}\mathfrak{D}(C\tensor{\mathcal{U}}{_A_B})\wedge\theta^B = \frac{1}{2} \tensor{\mathcal{S}}{_A_B_C}\theta^B\wedge\theta^C~.
            \end{equation}
            }
            \begin{equation}\label{eq:ZABC}
                \tensor{\mathcal{Z}}{_A_B_C} =\tensor{(\mathcal{U}^{-1})}{^D_{[B}} \eta_{A]E}\tensor{\mathcal{S}}{^{E}_C_D} + \tensor{\mathcal{U}}{^D_C}\tensor{(\mathcal{U}^{-1})}{^E_{[A}}\tensor{(\mathcal{U}^{-1})}{^F_{B]}}\eta_{DG}\tensor{\mathcal{S}}{^G_E_F}\,,
            \end{equation}
            where we defined
             \begin{align}
            \tensor{S}{^A_B_C} = - 2 \nabla_{[C} C \tensor{\delta}{_B_]^A} - 2 \nabla_{[C}( C \tensor{L}{^A_B_]} )\,.
        \end{align}
        This result is in agreement with Ref.~\cite{auth:EzquiagaFieldRedefinitions} in the case of $\tensor{L}{^A_B}= D(\phi, X) \nabla^A \phi \nabla_B \phi$. Using Eqs.~\eqref{eq:CartanTranPost} and \eqref{eq:ZABC}, the spin connection coefficient 
        are given by
        \begin{align}
                C\,\tensor{\tilde{\omega}}{_A_B_C} &=  \tensor{(\mathcal{U}^{-1})}{^D_C}\left(\tensor{\omega}{_A_B_D} +\tensor{\mathcal{Z}}{_A_B_D} \right)\nonumber \\
                &= \tensor{({\cal U}^{-1})}{^D_C}\,\tensor{\omega}{_A_B_D} - 2 \tensor{({\cal U}^{-1})}{^D_[_A}\eta_{B]C}  \nabla_D C\nonumber\\&\quad+ \left( \tensor{\delta}{_C^L}\tensor{\delta}{_{[A}^G}\tensor{\delta}{_{B]}^E}
                - \tensor{\delta}{_{[A}^L}\tensor{\delta}{_C^G}\tensor{\delta}{_{B]}^E} - \tensor{\delta}{_{[A}^L}\tensor{\delta}{_{B]}^G}\tensor{\delta}{_C^E} \right)\tensor{(\mathcal{U}^{-1})}{^D_L}\tensor{(\mathcal{U}^{-1})}{^F_G}\,\eta_{EI}\nabla_D\tensor{\mathcal{U}}{^I_F}\,.         \label{eq:Transformed_Cartan_Component}
    \end{align}
    
    An interesting aspect of the transformation \eqref{eq:Transformed_Cartan_Component} is the presence of a totally antisymmetric term explicitly separated from $\omega_{ABD}$, namely
    \begin{align}
     M^D=\frac{C}{4}\epsilon^{ABCD}\left(\tensor{\tilde{\omega}}{_A_B_C}-\tensor{(\mathcal{U}^{-1})}{^D_C}\tensor{\omega}{_A_B_D}\right)=\frac{1}{4}\epsilon^{DEFG}\tensor{(\mathcal{U}^{-1})}{^I_E}\tensor{(\mathcal{U}^{-1})}{^J_F}\nabla_J\tensor{\mathcal{U}}{_G_I}\,,\label{eq:Definition_M_A}
    \end{align}
    where we introduced the factor of $\frac{1}{4}$ for later convenience. It is precisely this term which later couples to the fermionic axial current, see Eq.~\eqref{eq:Final_Hermitian_Lagrangian}.

    \subsection{Disformal Dirac action}

    Let us now turn to the explicit transformation of the Dirac action. We start from the Dirac action in the hermitian formulation, which is given by
        \begin{equation}\label{eq:sdirac}
               S = \int \md^4x\, \sqrt{-\tilde g}  \Big[  \frac{i}{2} (\overline{\tilde \Psi} \gamma^B \tilde{\overset{\leftrightarrow}{\nabla}}_B \tilde \Psi) - \overline{\tilde \Psi} m \tilde \Psi\Big] \,,
            \end{equation}
        where, as in Sec.~\ref{sec:preliminaries}, the tilde denotes that the fermions are minimally coupled to the disformal metric $\tilde g_{\mu\nu}$. Expanding the hermitian derivative, that is
        \begin{align}\label{eq:covarianthermitian}
            \overline{\tilde \Psi} \gamma^A \tilde{\overset{\leftrightarrow}{\nabla}}_A \tilde \Psi =  - \partial_A \overline{\tilde \Psi} \gamma^A \tilde \Psi +\overline{\tilde \Psi } \gamma^A \partial_A \tilde \Psi + \frac{i}{2}  \overline{\tilde \Psi} \{\gamma^A,\Sigma^{BC}\} \tensor{\tilde \omega}{_A_B_C} \tilde \Psi\,,
        \end{align}
        showcases that the transformation amounts to computing $\tensor{\epsilon}{_D^A^B^C}\tensor{\tilde \omega}{_A_B_C}$. Using the following identity, namely
        \begin{align}
        \label{eq:identitysigma}
            \{\gamma^A, \Sigma^{BC} \} = \tensor{\epsilon}{_D^A^B^C}  \gamma^D \gamma^5\,,
        \end{align}
        and redefining the spinor field as
        \begin{equation}
              \Psi = C^{3/2} \sqrt{\textrm{det}( \mathcal{U} )}\,\tilde  \Psi\,,
        \end{equation}
        we obtain that the transformed Dirac action reads \begin{align}\label{eq:Final_Hermitian_Lagrangian}
                S = \int \md^4x\, \sqrt{-g} \Big[ \frac{i}{2} \tensor{(\mathcal{U}^{-1})}{^A_B} (\bar \Psi \gamma^B \overset{\leftrightarrow}{\nabla}_A \Psi) -  \bar \Psi C m \Psi - M_A J^A_5   \Big] \,,
            \end{align}
        where $M_A$ is defined in Eq.~\eqref{eq:Definition_M_A} and $J_5^A=\overline \Psi  \gamma^A\gamma^5  \Psi$. In the limit $L\to 0$, the action~\eqref{eq:Final_Hermitian_Lagrangian} reduces to the conformal case, see, e.g., Ref.~\cite{Quiros:2025ldk}.

        Note that, unlike the case of bosonic theories, the action of disformal transformations on the Dirac action decouples the conformal and disformal parts. The latter shifts the kinetic term 
        and introduces an additional axial term. Such coupling to the axial current stems from the non-holonomicity of the basis tetrads,  although it has a similar structure to that of torsion \cite{BeltranJimenez:2018vdo,BeltranJimenez:2019esp}. One may be tempted to interpret it as an effective non-metricity, but in this case, there would be no coupling to the axial current \cite{BeltranJimenez:2018vdo,BeltranJimenez:2019esp}. 
    
        To understand the physical interpretation of such a term, it is useful to consider the case of a vector disformal transformation, where $\tensor{\mathcal{U}}{^A_B}=\tensor{\delta}{^A_B}+D B^{A}B_{B}$. If we identify $B_A$ as being proportional to a fluid's 4-velocity, we see that $M^k$ is proportional to vorticity, namely $M^k\sim B^2\epsilon^{k0ij}B_0\nabla_iB_j\propto (\vec{\nabla}\times \vec{B})^k$. Thus, the origin of Eq.~\eqref{eq:Definition_M_A} is a spacetime vorticity induced by the disformal transformation of the tetrads. Note that such a coupling also appears in non-relativistic fermions in a rotating fluid, called the spin-vorticity coupling \cite{PhysRevB.96.020401}.

\subsubsection{Comparison with SME coefficients}

To facilitate comparison of Eq.~\eqref{eq:Final_Hermitian_Lagrangian} with the SME coefficients $c_{\mu\nu}, b_\mu$ introduced in Eq.~\eqref{GammaM}, we explicitly separate $\tensor{(\mathcal{U}^{-1})}{^A_B}$, given by Eq.~\eqref{eq:ThetaDisformalInv}, into a trace and traceless part, namely
\begin{align}
  \label{eq:tracelessP}\tensor{(\mathcal{U}^{-1})}{^A_B}=\left(1+\tfrac{1}{4}{\rm Tr}P\right)\left(\tensor{\delta}{^A_B}+\tensor{\mathcal{P}}{^A_B}\right)\quad{\rm with}\quad \tensor{\mathcal{P}}{^A_B}=\frac{1}{\left(1+\tfrac{1}{4}{\rm Tr}P\right)}\left(\tensor{{P}}{^A_B}-\frac{1}{4}{\rm Tr}P\,\tensor{{\delta}}{^A_B}\right)\,.
\end{align}
By definition ${\rm Tr}{\cal P}=0$.
We then canonically normalize the kinetic term of the fermions via an additional field redefinition given by $\Psi\to \left(1+\tfrac{1}{4}{\rm Tr}P\right)^{-1/2}\Psi$. After the field redefinition, the action becomes
\begin{align}\label{eq:tracelessdirac}
S = \int \md^4x\, \sqrt{-g} \Big[ \frac{i}{2} \bar \Psi \gamma^A \overset{\leftrightarrow}{\nabla}_A \Psi -  \bar \Psi m_\Psi \Psi +\frac{i}{2} \tensor{{\cal P}}{^A_B} \bar \Psi \gamma^B \overset{\leftrightarrow}{\nabla}_A \Psi - {\cal M}_A J^A_5   \Big] \,,
\end{align}
where, for compactness, we have defined
\begin{align}\label{eq:ms}
m_\Psi=\frac{C m}{1+\tfrac{1}{4}{\rm Tr}P}\quad{\rm and}\quad {\cal M}_A=\frac{M_A}{1+\tfrac{1}{4}{\rm Tr}P}\,.
\end{align}
Note that we could have defined $L$ or $P$ to be traceless from the start by shifting the conformal factor, but we considered it clearer to perform the transformation at the end of the calculation.
Comparing Eq.~\eqref{eq:tracelessdirac} with the fermion SME action Eq.~\eqref{LSME}, we identify the correspondence
\begin{align}
\label{cbPM}
& \tensor{c}{_{\mu\nu}} \leftrightarrow -\tensor{{\cal P}}{_{\mu\nu}}, \\
& \hspace{1.75mm} b_\mu \leftrightarrow -{\cal M}_\mu\,, \nonumber
\end{align}
where ${\cal P}_{\mu\nu}$ and ${\cal M}_\mu$ are given by Eqs.~\eqref{eq:tracelessP} and \eqref{eq:ms}, respectively, generalizing the special case given in Eq.~\eqref{eq:correspondence}. Note that we employ the action Eq.~\eqref{eq:tracelessdirac} and the correspondence~\eqref{cbPM} in Sec.~\ref{sec:applications} and Apps.~\ref{app:isolorentzviolations} and \ref{app:boostviolations}. In the rest of the paper, we work with Eq.~\eqref{eq:Final_Hermitian_Lagrangian} for convenience.

    \section{Single Field Disformal Transformation}
    \label{sec:Examples}

    Let us now focus on the cases of scalar and vector disformal transformations. We discuss the inclusion of the field strength in the disformal part in App.~\ref{sec:U_!_coupling}. In a subsequent publication \cite{auth:us2}, we derive the conditions on disformal transformations for consistent couplings to fermions, ranging from multi-scalar fields to higher-order derivatives.

    \subsection{Scalar Field}
    \label{ssec:scalarfield}
    Making use of both Eqs.~\eqref{eq:ThetaDisformal} and \eqref{eq:ThetaDisformalInv}, the scalar disformal transformation is expressed as
    \begin{align}
        \tensor{L}{^A_B} = D  \nabla^A \phi \nabla_B \phi, \qquad \tensor{P}{^A_B} = - \frac{D}{1 + D X} \nabla^A \phi \nabla_B \phi \,,
    \end{align}
    where $D=D(\phi,X)$, $C=C(\phi,X)$ and $X= \nabla_\mu \phi \nabla^\mu \phi$. Since $\nabla_{[A} \nabla_{B]} \phi =0$ the vorticity term \eqref{eq:Definition_M_A} drops out (that is $M_A =0$), leading to  
    \begin{align}
    \label{Sscalar}
        S= &  \int \md^4x\,\sqrt{-g} \Big[ \frac{i}{2}    \bar \Psi \gamma^A \overset{\leftrightarrow}{\nabla}_A \Psi -  \frac{i}{2} \frac{D}{1+ D X} \nabla^A \phi \nabla_B \phi  \,\bar \Psi \gamma^B \overset{\leftrightarrow}{\nabla}_A \Psi   -  \bar \Psi C m \Psi  \Big]  \,.
    \end{align}
    
    For easier comparison with the literature, consider the case of a massless fermion, discussed in Refs.~\cite{Brax:2012hm, Brax:2014vva,Brax:2015hma}, in which case the action \eqref{Sscalar} simplifies to
    \begin{align}
    \label{eq:diracmzero}
        S= &  \int \md^4x\,\sqrt{-g} \Big[ \frac{i}{2}    \bar \Psi \gamma^A \overset{\leftrightarrow}{\nabla}_A \Psi -   \frac{D}{1+ D X} \nabla_\mu \phi \nabla_\nu\phi T^{\mu\nu}   \Big]  \,,
    \end{align}
    where we used
    \begin{equation}
                \tensor{T}{_\mu_\nu} = \frac{i}{4}\bar{\Psi}\left[ \gamma_\mu \overset{\leftrightarrow}{\nabla}_\nu + \gamma_\nu \overset{\leftrightarrow}{\nabla}_\mu\right]\Psi\,.
     \end{equation}
     Note that the disformal coupling in Eq.~\eqref{eq:diracmzero} is also valid for non-vanishing mass, when evaluated on-shell \cite{Brax:2014vva}. 

    Our result agrees with Refs.~\cite{Brax:2012hm, Brax:2014vva,Brax:2015hma} at leading order in the derivative expansion, that is, for $DX\ll1$. However, it is more general, since we do not rely on the derivative expansion. In particular, it is often assumed that the disformal coupling appearing in the metric, that is ${\cal D}$ in Eq.~\eqref{eq:disformalpreliminary}, is constant. But fermions couple to the tetrad and, hence, to $D$.
     Interestingly, a constant ${\cal D}$ does not necessarily imply a constant $D$. In fact, $D$ depends on ${\cal D}$ and $X$ through $D=X^{-1}(-1 +\sqrt{1+{\cal D}X})$. Thus, in general, a constant ${\cal D}$ yields an $X$-dependent $D$.

    \subsection{Vector Field}
     \label{ssec:vectorfield}

    In the case of the vector disformal transformation, $L$ and $P$ take a similar form, that is
    \begin{align}
     \label{LPsingle}
            \tensor{L}{^A_B} = D B^A B_B, \qquad \tensor{P}{^A_B} = -\frac{D}{1+ D X} B^{A} B_B\,,
    \end{align}
    where $D=D( X ) $, $C=C(X )$ and $X= B_\mu B^{\mu}$. In contrast to the scalar case, substituting into Eq. \eqref{eq:Definition_M_A} results in an axial coupling given by
        \begin{align}
            M_A = - \frac{ D^2 X }{8  (1 + D X)} \tensor{\epsilon}{_A^B^C^D} B_D {\cal F}_{BC}= - \frac{ D^2 X }{4  (1+ D X)}  B^D  {\cal F}^{\star}_{AD}\,,
            \label{eq:MAD}
        \end{align}
    where ${\cal F}_{BC}$ is the field strength of the vector field, and $ {\cal F}_{AD}^{\star}$ its Hodge dual. Such a term has appeared before in the literature Ref.~\cite{Bittencourt:2015ypa}, where the authors performed a disformal transformation of the Dirac equation. Note that to have a non-vanishing contribution from the axial coupling in cosmology, the vector field has to be inhomogeneous, and, therefore, such interactions should be suppressed on cosmological scales.

        We end this section by explicitly writing the Dirac action after the transformation, which reads
        \begin{align}
        \label{Svector}
            S =   \int d^4 x\sqrt{-g} \Big[ \frac{i}{2}  \bar \Psi \gamma^A \overset{\leftrightarrow}{\nabla}_A \Psi -  \bar \Psi C m \Psi-  \frac{i}{2} \frac{D}{1+ D X} B^{ A} B_B \, \bar \Psi \gamma^B \overset{\leftrightarrow}{\nabla}_A \Psi   +   \frac{ D^2 X }{4  (1+ D X)}  B^{D}  {\cal F}^{\star}_{AD} J_5^A \Big] \,.
        \end{align}
Note that our new contribution, that is, the coupling to the axial current, only appears at higher orders in the derivative expansion. 
Also note that since the SME coefficient $b_\mu$ is only generated for vector fields and has a suppressed quadratic dependence on $D$, we devote attention to the $c_{\mu\nu}$ coefficients for the remainder of the paper.

\section{Application}\label{sec:applications}
We are now positioned to focus on a concrete scenario of the single field formalism (Sec.~\ref{sec:Examples}): scalar and vector ULDM with quadratic couplings to QED. Quadratic couplings appear naturally in disformal transformations\footnote{See also Ref.~\cite{Jiang:2024agx} for a motivation of quadratic couplings based on a $\mathbbm{Z}_2$ symmetric theory.} and
are of particular phenomenological interest because they address DM while avoiding the stringent constraints imposed by fifth-force experiments~\cite{Brax:2014vva}. Instead, they give rise to apparent Lorentz-violating effects, which can be mapped onto SME operators. As the disformal couplings are quadratic in the fields, our scenario also applies to pseudoscalar ULDM. Both cases of scalar and vector ULDM are accounted for in the action Eq.~\eqref{eq:tracelessdirac}, complemented by the standard kinetic terms for a scalar and vector of mass $m_\phi$ and $m_B$, respectively. 

Since we are interested in gravitational couplings, we neglect potential kinetic-mixing effects~\cite{Holdom:1985ag,Fabbrichesi:2020wbt} which are the focus of most dark photon searches and strongly constrained in the case of vector ULDM (see e.g. horizontal branch bounds in Ref.~\cite{Vogel:2013raa}). Specifically, we omit the term $\propto F_{\mu\nu}\mathcal{F}^{\mu\nu}$.

    \subsection{Ultralight Dark Matter}
    \label{ssec:ULDM}

Ultralight dark matter is an umbrella term describing DM candidates within the approximate mass range $10^{-22}$\;eV $\lesssim m \lesssim  1$\;eV. 
The lower end, $m\sim 10^{-22}~$eV, assumes ULDM accounts for all of the DM and is primarily constrained by galactic structure formation, while the upper end is set by requiring that the DM occupation number per de Broglie wavelength in galaxies remains large compared to unity, implying that ULDM candidates are bosonic.
Present-day DM is cold and nonrelativistic, with a virial velocity $|\vec{v}|\sim 10^{-3}$ in the Milky Way. Ultralight DM is temporally coherent over $\tau_c \approx 4\pi/(m |\vec{v}|^2)\sim 10^{6} \lambdabar_{C}$ and spatially coherent over $l_c \approx \lambdabar_{\rm dB} =2\pi/(m |\vec{v}|) = 10^{3}\lambdabar_{C}$, where $\lambdabar_{\rm dB}$ and $\lambdabar_{C}$ are the reduced de Broglie and Compton wavelengths, respectively. For the masses of interest, we have that $\lambdabar_{C}\sim 10^6\,{\rm km}\,(10^{-15}{\rm eV}/m)\sim 4 \,\mu{\rm s}\,(10^{-15}{\rm eV}/m)$. For comparison, one astronomical unit is $1 {\rm AU}\sim 10^8\,{\rm km}$.
This allows a classical wavelike description in which the particle mass predominantly controls the macroscopic dynamics. 

One of the simplest and well-motivated models of ULDM is the axion, or more broadly, axion-like particles. By virtue of the misalignment mechanism~\cite{Preskill:1982cy,Abbott:1982af,Dine:1992vx}, the axion admits a wavelike description and could account for all of the DM in the universe.
More broadly, a generic scalar or pseudoscalar field coupled to the SM can be described with an EFT-based approach with
additional
higher-dimensional 
($d\geq 5$)
operators normalized to the Planck scale.
Several of the so-called linear ($d=5$) couplings for masses below $m\lesssim 10^{-6}$~eV are highly constrained by tests of the Equivalence Principle~\cite{Schlamminger:2007ht,Smith:1999cr,MICROSCOPE:2022doy,Wagner:2012ui} and fifth-force searches~\cite{Adelberger:2003zx,Fischbach:1996eq,Lee:2020zjt,Tan:2020vpf} (see also, e.g., Ref.~\cite{Will:2014kxa}).
This has, in part, motivated broad searches for
quadratic ($d=6$) operators~\cite{Banerjee:2022sqg}.
For a comprehensive and recent overview of ULDM, see, e.g., Ref.~\cite{Eberhardt:2025caq}; for a review of dark photons, see, e.g., Ref.~\cite{Caputo:2021eaa}.

Special phenomenological considerations aside, the solutions for a classical, non-relativistic (and thus massive) scalar $\phi$ or vector $B^\mu$ ULDM field are
\begin{align}
&\phi(t,\vec{x}) =\frac{\sqrt{2\rho_{\rm DM}}}{m_\phi}\cos(\omega_\phi t-\vec{k}\cdot \vec{x}), \label{scalarULDM}\\
&B_\mu(t,\vec{x}) = \frac{\sqrt{2\rho_{\rm DM}}}{m_B}\cos(\omega_B t-\vec{k}\cdot \vec{x})\epsilon_\mu\,, \label{vectorULDM}
\end{align}
where $\epsilon_\mu$ is a real polarisation vector (which may include a longitudinal mode since the vector is massive) and is, in principle, independent of $\vec{k}\approx m\vec{v}$ (see, e.g., vector ULDM searches with interferometers \cite{KAGRA:2024ipf}). In Eqs.~\eqref{scalarULDM} and \eqref{vectorULDM}, we take $\omega \approx m\sqrt{1+|\vec{v}|^2}$ and  $|\vec{k}|\approx m|\vec{v}| \ll m$, where the DM wind velocity  $|\vec{v}|\sim 10^{-3}$ accounts for the boost from the galactic DM rest frame to a solar-system frame. For simplicity, we consider plane waves only. 
Note that in this limit the divergence of the Proca equation, $\partial_\mu B^\mu =0$, implies that the timelike component of the vector field $B^0 \approx \vec{v}\cdot \vec{B} \ll |\vec{B}|$.
The amplitude $\sqrt{2\rho_{\rm DM}}/m$ is set by equating the time-averaged energy density to the local DM density $\rho_{\rm DM} \approx 0.3$\;GeV/cm$^3$.

Before going into the details on the bounds, let us qualitatively discuss possible spacetime properties of the scalar and vector fields, ultimately linked to their generation mechanism (see e.g. Refs.~\cite{Caputo:2021eaa,Cyncynates:2024yxm}). First, a scalar field is, as very often in cosmology, mostly time-dependent with small spatial inhomogeneities. This means that any Lorentz-violating effect from the disformal coupling is mainly time-dependent and mostly isotropic. Second, the vector has two distinct possibilities due to the polarization. The first possibility is that the vector points in the same direction throughout the solar system; this is called the "\textit{fixed polarization scenario}" in Ref.~\cite{Caputo:2021eaa}.\footnote{There is also a "\textit{random polarization scenario}" which assumes that the vector points in a random direction after every coherence length \cite{Caputo:2021eaa}. This is somewhat similar to the \text{fixed polarization scenario} and, therefore, we will not consider it further.} The second, opposite possibility is that the vector is unpolarized, and there results no preferred spatial direction after one coherence length. We will consider these two extreme cases for the vector to derive optimistic and conservative estimates for the disformal coupling to vectors. We proceed now to discuss the bounds.

\subsection{Bounds on Disformal Couplings from Searches for Apparent Lorentz Violation}
\label{ssec:constraints}
Since the disformal transformation couples universally to fermions, we can exploit results from studies on the coefficients $c^f_{\mu\nu}$ of any fermion species $f$. 
Broadly speaking, searches for Lorentz violation can be grouped into searches for isotropic Lorentz-violating effects and searches for a preferred direction in spacetime. 
As their name suggests, isotropic Lorentz-violating effects arise from rotationally invariant combinations of controlling coefficients. These modify dispersion relations in an orientation-independent way and can be constrained by searches for vacuum \v{C}erenkov radiation or photon decay~\cite{PhysRevD.1.961,Altschul:2006zz,Coleman:1997xq,Lehnert:2004be,Hohensee:2008xz,Klinkhamer:2008ky}, among others. Three of the nine physical components of $c_{\mu\nu}$ contribute to this category of effects.

In contrast, controlling coefficients inducing a preferred direction in spacetime lead to anisotropic effects, which, in Earth-based laboratories, translate into time-varying signals determined by the Earth's axial rotation and revolution around the Sun. 
For this reason, it has become standard practice to report constraints on controlling coefficients in the Sun-centered frame (SCF)~\cite{Bluhm:2001rw,Kostelecky:2002hh,Bluhm:2003un}, which remains essentially fixed and nonrotating over the duration of most experiments. In this frame, the $Z$ axis is parallel to the Earth’s axis of rotation, the $X$ axis points from the Earth toward the Sun at the 2000 vernal equinox, and the $X$-$Y$ plane is parallel to the equatorial plane. 
The remaining six components of the $c_{\mu\nu}$ coefficients belong to this category. From now on, we focus on anisotropic Lorentz-violating effects, since they lead to the stricter constraints. We discuss bounds from isotropic Lorentz-violation in App.~\ref{app:isolorentzviolations} and from complementary Lorentz boost violations in App.~
\ref{app:boostviolations}.

According to the SME data tables~\cite{auth:KosteleckTables2025}, the most stringent bounds on the $c_{\mu\nu}$ coefficients are placed by a search for time-dependent signatures in a $^{21}$Ne-Rb-K comagnetometer experiment~\cite{Smiciklas:2011xq}. 
The experiment~\cite{Smiciklas:2011xq} used a search for time-varying signals in a $^{21}$Ne-Rb-K comagnetometer experiment to constrain the neutron-sector SME coefficients $c^n_{\mu\nu}$.\footnote{The SME is typically viewed as constructed via fundamental particle fields. Here, an analogous EFT description at the nucleon level is assumed, from which the standard approach of taking the nonrelativistic limit follows.} 
Two of the six components inducing time-dependent signatures do not appear in the observable considered, as the nonrelativistic perturbed Hamiltonian depends only on the spatial components $c^n_{ij}$. The remaining four components $c^n_{YZ}, c^n_{XZ}, c^n_{XY}$ and $c^n_{XX}-c^n_{YY}$ can be viewed as parametrizing violations of the $SO(3)$ rotation subgroup of the Lorentz group.
The quoted results are~\cite{Smiciklas:2011xq}:
\begin{align}
\label{neutronlimits}
\begin{split}
c^n_{YZ} + c^n_{ZY} &< (4.8\pm 4.4)\times 10^{-29}, \\
c^n_{XZ} + c^n_{ZX} &< (-2.8\pm 3.4)\times 10^{-29},\\
c^n_{XY} + c^n_{YX} &< (-1.2\pm 1.4)\times 10^{-29}, \\
c^n_{XX} - c^n_{YY} &< (1.4\pm 1.7)\times 10^{-29}. 
\end{split}
\end{align}

Indeed, let us discuss under which conditions the sets of results in Eq.~\eqref{neutronlimits} (and also later in Eq.~\eqref{eq:BASElims}), may be converted into constraints on the disformal parameter $D$. 
The hypothesis tested by Refs.~\cite{Smiciklas:2011xq,Wursten:2023bsl} is that of a flat spacetime metric and spacetime-position-independent $c^n_{\mu\nu}$ coefficients in the SCF. These are reasonable assumptions given the weak local gravitational field and the smallness of the SME coefficients~\cite{Kostelecky:2003fs}. They also fix the disformal coupling $D$ to be constant. 
Then, to a good approximation, the SCF coefficients are related to those in the non-inertial laboratory frame by a time-dependent rotation involving the laboratory colatitude and harmonics of the Earth's sidereal frequency~\cite{Bluhm:2001rw,Kostelecky:2002hh,Bluhm:2003un,Ding:2020aew}.

The comagnetometer signal was measured every 22 seconds (s), modulo periodic pauses of 300-400 s to calibrate the experiment, but otherwise regularly over a three-month period~\cite{Smiciklas:2011xq}. 
The latter total duration of $T_{\rm exp} = 3$ months and the minimum measurement interval $T_{\rm min} = 22$\;s set the relevant temporal scales. 
If the ULDM field only varies on timescales greatly exceeding $T_{\rm exp}$, its amplitude and direction are essentially constant and can directly be compared to the SME bounds. This is only the case for ULDM masses $m \lesssim 10^{-22}$\;eV, which are nominally outside of the ULDM range (though in the rather restrictive case that a single candidate accounts for all DM).
Consequently, we focus on ULDM masses $m \gtrsim 10^{-16}$\;eV, in which case the oscillation period is shorter than the minimum measurement interval $T_{\rm min}$ and the oscillations average out between measurements. At the lower end of this mass range, the temporal coherence $\tau_c\approx 6T_{\rm exp}$, while moving towards larger masses necessitates the accounting for stochastic amplitude fluctuation, which degrades the signal by an ${\cal O}(1)$ factor (see, e.g., Ref.~\cite{Centers:2019dyn}). 
In order to estimate the sensitivity of the aforementioned experiments to ULDM, we thus time-average the objects identified in Eq.~\eqref{cbPM}, and assume a classical, non-relativistic background field with a mass $m$ and $m v_\mu \sim (\omega, -\vec{k})\sim m(1,-\vec{v}) $, where $v = |\vec{v}| \sim  10^{-3}$ corresponds to the velocity of the Sun with respect to the DM background.

\subsubsection{Scalar ULDM}
For the single scalar ULDM scenario, which was also considered in Ref.~\cite{Jiang:2024agx}, $\langle X \rangle = \langle \partial_\mu \phi \partial^\mu \phi \rangle 
=\rho_{\rm DM}$, 
where $\langle X \rangle$ refers to oscillation average. Note that $\langle X \rangle$ is positive for a mostly time-dependent scalar field, that is $\dot\phi\gg|\nabla\phi(t)|$. 
For the $c_{\mu\nu}$ tensor we find~\footnote{Formally this result is independent of $f$ and matches to Eq. (3.19) of Ref.~\cite{Jiang:2024agx}, where ${\cal C}_2  \to\rho_{\rm DM} D / [1+(3/4)\rho_{\rm DM} D]$.} 
\begin{equation}
\label{eq:c_scalar}
\langle c^n_{\mu\nu}\rangle = \frac{D }{1+\frac{3}{4} D \langle X \rangle}\left\langle \partial_\mu \phi \partial_\nu \phi -\frac{1}{4}\eta_{\mu\nu}X\right\rangle= \frac{\rho_{\rm DM}D}{1+\frac{3}{4}\rho_{\rm DM}D}\left(
v_\mu v_\nu - \tfrac{1}{4}\eta_{\mu\nu}\right)\,,
\end{equation}
where recall that $v_\alpha v^\alpha=1$.
 The velocity suppression $|\vec{v}|^2\sim 10^{-6}$, which enters via $v_\mu \approx (1,\vec{v})$ and arises from the weak spatial dependence of Eqs.~\eqref{scalarULDM} and \eqref{vectorULDM}, significantly reduces the sensitivity to the disformal coupling $D$ via the off-diagonal $c^n_{\mu\nu}$ components (or linear combinations like $c^n_{XX}-c^n_{YY}$ where the $\eta_{\mu\nu}$ pieces drop out). 
 As a result, the stringent comagnetometer constraints in Eq.~\eqref{neutronlimits} have a reduced impact on $D$. 
With $\rho_{\rm DM} \approx 0.3$\;GeV/cm$^3 \approx 2\times 10^{-6}$\;eV$^4$ and $[D] = $\;GeV$^{-4}$, and assuming the components of $\vec{v}$ are of similar magnitude, the constraint on the $YZ$ combination translates into the upper bounds
\begin{align}
\label{scalarDLambdaconstraints}
\begin{split}
&D \lesssim 2\times 10^{19}\;\text{GeV}^{-4}\\
&\Lambda \gtrsim 15\;\text{keV}    
\end{split}\quad \text{(scalar ULDM)}\,,
\end{align}
where $\Lambda$ is the energy scale associated with $D$, namely  $\Lambda\equiv D^{-1/4}$. The sensitivity to $\Lambda$ is slightly better than found in Ref.~\cite{Jiang:2024agx}, which instead focused on atomic-clock experiments that are primarily sensitive to the electron $c^e_{\mu\nu}$ coefficients constrained at the $\sim 10^{-21}$ level. Note that complementary test using monophoton searches at CMS~\cite{CMS:2012lmn,CMS:2014rwa} have led to roughly ${\cal O}(10^6)$ stronger constraints $\Lambda \gtrsim 490~$GeV, as they are not suppressed by the tiny factor $\rho_{\rm DM}|\vec{v}|^2 \sim 10^{-48}~\mathrm{GeV}^4$~\cite{Brax:2014vva}.

\subsubsection{Vector ULDM \label{subsub:vectoruldm}}
We now consider the vector ULDM field $B^\mu$ (cf. Eq.~\eqref{vectorULDM}). Recall that in this case $[D] = $\;GeV$^{-2}$ and that the component $B^0$ is velocity suppressed (see discussion in Sec.~\ref{ssec:ULDM}). 
Our considerations broadly mirror the scalar-field case, with a few key distinctions. 
First, it is clear from the mass dimensions that the sensitivity to vector ULDM will depend quadratically on the mass $m_B$ of the dark vector; indeed, the equivalent of Eq.~\eqref{eq:c_scalar} for dark vectors is
\begin{align}
    \langle c_{\mu\nu}^n \rangle= \frac{D }{1+\frac{3}{4} D \langle X \rangle}\left\langle B_\mu B_\nu -\frac{1}{4}\eta_{\mu\nu}X\right\rangle
    = \frac{\rho_{\rm DM} D}{m_B^2 -\frac{3}{4} \rho_{\rm DM}D}\left(\epsilon_\mu \epsilon_\nu + \frac{1}{4}\eta_{\mu\nu} \right)\,,
    \label{eq:c_vec_general}
\end{align}
where we used that $\epsilon_\alpha \epsilon^\alpha=-1$.
We can see from this expression that, for a fixed bound on $c_{\mu\nu}^n$, the sensitivity to the disformal coupling $D$ will increase quadratically towards lower vector masses $m_B$. We recall that in the scalar field case, this dependence cancels out in $\partial_\mu \phi$. 
Secondly, in the vector case, it is the polarisation vector, $\epsilon_\mu$, rather than the field gradient, $\partial_\mu \phi$, that selects an orientation in the SCF. 
Consequently, the constraints on the disformal coupling $D$ depend both on our assumptions about the mass $m_B$ and the polarisation $\epsilon_\mu$ of the vector field.
Unfortunately, it is currently impossible to make a precise statement about the polarisation distribution of a potential vector ULDM background (see also Ref.~\cite{Caputo:2021eaa}). In the following, we will consider the case when the vector field points in a fixed direction since this is widely assumed in searches \cite{Caputo:2021eaa}. We discuss the case of unpolarized vectors with a net isotropic effect in App.~\ref{app:isolorentzviolations} but report the corresponding bounds at the end of this section.

If the polarisation $\epsilon_\mu$ is fixed across the patch of spacetime probed by experiment, it follows that $\langle X \rangle = - \rho_{\rm DM}/m_B^2$. In principle, the direction $\epsilon_\mu$ is arbitrary. However, for illustration purposes, we consider the case of $\epsilon_\mu = \frac{1}{\sqrt{3}}(0,1,1,1)_\mu$, in the SCF, where we neglected terms proportional to $|\vec{v}|$. Different directions will differ by ${\cal O}(1)$ factors. From Eq.~\eqref{eq:c_vec_general},  we obtain the following $c^n_{\mu\nu}$  in the SCF at leading order in $|\vec{v}|$:
\begin{align}
\label{cvectorULDM}
c^n_{TT} =  \frac{1}{4}\frac{\rho_{\rm DM}D}{m_B^2-\tfrac{3}{4}\rho_{\rm DM}D}\,,\qquad
c^n_{JK} =  \frac{\rho_{\rm DM}D}{m_B^2-\tfrac{3}{4}\rho_{\rm DM}D}\left(\frac{1}{3} - \frac{1}{4}\delta_{JK}\right)\,,\quad J,K \in \{X,Y,Z\}\,,
\end{align}
or, in matrix form,
\begin{align}
    \big[c^n_{\mu\nu}\big] =\frac{\rho_{\rm DM}D}{m_B^2-\tfrac{3}{4}\rho_{\rm DM}D}\begin{pmatrix}
        \frac{1}{4} & 0 & 0 & 0\\
        0 & \frac{1}{12} & \frac{1}{3} & \frac{1}{3}\\
        0 & \frac{1}{3} & \frac{1}{12} & \frac{1}{3}\\
        0 & \frac{1}{3} & \frac{1}{3} & \frac{1}{12}
    \end{pmatrix}\,.
\end{align}

As a consequence of the lower mass dimension of $X=B_\mu B^\mu$ in the vector case versus $X=\partial_\mu \phi \partial^\mu \phi$ in the scalar case, the translation of the SME into bounds on the disformal coupling of the vector, Eq.~\eqref{eq:c_vec_general}, depends explicitly on the mass of the vector field, while Eq.~\eqref{eq:c_scalar} did not. Note, however, that the bounds of applicability of Eq.~\eqref{eq:c_scalar} and \eqref{eq:c_vec_general} are similar since both formulas assume that the experiment averages over the background field oscillation. 
While noting various model-dependent caveats, we argue that
the mass scale of $m = 10^{-15}$\;eV is a sensible choice to illustrate the constraints. For this mass scale, the coherence time is comparable to the observation time ($\tau_c \approx T_{\rm exp}$) and the coherence length is $l_c \approx 8$ AU.

We obtain for the fixed polarization scenario that
\begin{align}
\label{vectorDLambdaconstraints}
\begin{split}
&D \lesssim 6\times 10^{-35}\;\text{GeV}^{-2}\left(\frac{m_B}{10^{-15}\,{\rm eV}}\right)^{2}\\
&\Lambda \gtrsim  10^{17}\;\text{GeV}\left(\frac{m_B}{10^{-15}\,{\rm eV}}\right)^{-1}    
\end{split}
\quad \text{(vector ULDM)}
\end{align}
where we defined $\Lambda\equiv D^{-1/2}$.
We note that in this scenario, the constraints listed in Eq.~\eqref{neutronlimits} probe close-to-Planckian $\Lambda$ scales. In the case of unpolarized vector ULDM where the net effect is isotropic Lorentz-violation, we find in App.~\ref{app:isolorentzviolations}, that the bounds become substantially weaker with $D\lesssim 10^{-20}\,{\rm GeV}$ and $\Lambda\gtrsim 10^{10},{\rm GeV}$, with the same mass dependence as in Eq.~\eqref{vectorDLambdaconstraints}. Note that they are still stringent as compared to the scalar field case, Eq.~\eqref{scalarDLambdaconstraints}.

Lastly, it should be noted that in addition to tests of rotation invariance, tests of boost invariance can also constrain $D$ and have an interesting connection to disformal couplings involving gauge fields. However, the constraints we find in this case are weaker than Eq.~\eqref{scalarDLambdaconstraints} and Eq.~\eqref{vectorDLambdaconstraints}. We refer the curious reader to a thorough investigation of these topics as provided in App.~\ref{app:boostviolations}.

\section{Discussion and conclusions \label{sec:conclusions}}

Searches for Lorentz-violating effects offer a window to test gravity theories with higher derivatives via disformal couplings to matter. Here, we have laid out the general formalism to explore disformal couplings to arbitrary fields without assuming any derivative expansion. We then provided a correspondence with the SME coefficients and applied it to the case of scalar and vector ULDM disformally coupled to SM fermions. For scalar ULDM, we find that current Lorentz violation searchers bound the energy scale associated with the disformal coupling to $\Lambda > 10\,{\rm keV}$. Note that the strongest constraints on disformal couplings to scalar dark matter come from particle colliders, setting $\Lambda > 100 \,{\rm GeV}$ \cite{Kaloper:2003yf,Brax:2014vva}. Although weaker, the bound presented in this work is a complementary probe to such a scenario.

For vector ULDM, the bounds from Lorentz violation searches are significantly stricter than in the scalar case. This is because, while $\dot\phi^2\sim \rho_{\rm DM}$, for a massive vector $B_\mu B^\mu \sim \rho_{\rm DM}/m_B^2$. Thus, for small masses, a quadratic coupling to $B_\mu B_\nu$ is enhanced by a factor $1/m_B^2$ with respect to $\partial_\mu\phi\partial_\nu\phi$. We then derived the first bounds on disformally coupled vector ULDM. We found that $\Lambda > 10^{17}\,{\rm GeV}$ in the case of "polarized" vector ULDM (by polarized, we mean that the vector points at a fixed direction within a coherence length). Such a stringent bound is relaxed if we consider the case of "unpolarized" vector ULDM (with a net isotropic Lorentz-violating effect after a coherence length). If so, the bound is reduced to $\Lambda > 10^{10}\,{\rm GeV}$. Note that even the conservative bound is eight orders of magnitude larger than the best bound on the disformal coupling to scalar ULDM.

It should be noted that it is still not clear what the resulting polarization of the ULDM vector would be after structure formation and virialization. It would be interesting to derive more accurate bounds for a given generation model and the subsequent nonlinear evolution. This is out of the scope of this paper. However, we do not expect that the order of magnitude of our conservative estimate will be significantly altered. It would also be interesting to apply our general formalism of disformal couplings to other fields and away from the derivative expansion. We leave these issues for future work.

We end this paper with a discussion on possible implications of the derived bounds on cosmology. The most interesting regime for cosmology is when the disformal couplings become of ${\cal O}(1)$, since this is when we expect significant modified gravity effects from higher derivatives (see, e.g., Ref.~\cite{Zumalacarregui:2013pma}). In that case, one cannot neglect the coupling to matter in the gravitational action. Equivalently, one may directly work with the gravitational metric $\tilde g_{\mu\nu}$ rather than $g_{\mu\nu}$, leaving SM particles minimally coupled. For example, in the disformal inflation model \cite{Kaloper:2003yf}, cosmic inflation happens for energy scales above that of the disformal coupling, that is $\Lambda$. Interestingly, at those high energies, higher derivatives render the gravitational metric $\tilde g_{\mu\nu}$ to be that of an exponentially expanding universe. Note that Ref.~\cite{Kaloper:2003yf} chooses $\Lambda \sim {\rm TeV}$ and results in (low-scale) inflation, but could be in principle higher.  

Let us estimate the times (or, equivalently, the energy scale of the Universe) when higher derivatives could be relevant for the gravitational metric, given our bounds on scalar and vector ULDM. To do so, we find the time when disformal couplings are large, namely when $DX\sim 1$. First, for scalar ULDM we have that $DX\approx D\rho_{\rm DM}=D\rho_{\rm DM,0}/a^3$, where $a$ is the scale factor of the Universe and $\rho_{\rm DM,0}\approx 10^{-47}{\rm GeV}^4$ is the mean dark matter energy density in the Universe today as given by Planck \cite{Planck:2018vyg}. Thus, modified gravity effects could be important for the expansion of the universe for $a<a^S_{\rm MG}\sim (D\rho_{\rm DM,0})^{1/3}$, where the subscript "MG" stands for modified gravity. For vector ULDM, we have that $X\approx -\rho_{\rm DM}/m_B^2$, and so $a<a^V_{\rm MG}\sim (D\rho_{\rm DM,0}/m_B^2)^{1/3}$. Using the constraints \eqref{scalarDLambdaconstraints} and \eqref{vectorDLambdaconstraints}, we respectively find $a^S_{\rm MG}\sim 6\times 10^{-10}$ and $a^V_{\rm MG}\sim 8\times 10^{-12}$ (the latter becomes $a^V_{\rm MG}\sim 10^{-7}$ for the conservative bound from App.~\ref{app:isolorentzviolations}). 

Note that the above derived time (scale factor), when disformal couplings effects could be important in the early universe's expansion history, corresponds to times much before Big Bang Nucleosynthesis (BBN), which roughly starts at around $a_{\rm BBN}\sim 10^{-9}$ (see Ref.~\cite{Cyburt:2015mya} for a review). Namely, we find that $a^{S,V}_{\rm MG}<a_{\rm BBN}$. A small caveat, though, is that for ULDM there will be a time below which the ULDM field is frozen by the cosmic expansion, corresponding to $H\gg m$. The time of friction domination is given by $a<a_m\approx 10^{-10}(m/10^{-15}\,{\rm eV})^{-1/2}$. Thus, for scalar ULDM, we conclude that disformal couplings will become large before the scalar ULDM field stops behaving like dark matter, that is $a_{\rm MG}>a_m$, for $m>6\times 10^{-17}\,{\rm eV}$. For the vector ULDM, we find that $a_{\rm MG}>a_m$ implies $m>3\times 10^{-13}\,{\rm eV}$ (or $m> 10^{-22}\,{\rm eV}$ for the conservative bound). For smaller masses, the scalar (vector) ULDM field becomes friction-dominated for $a<a_m$, and the current bound from backreaction does not apply. It would be interesting, nevertheless, to investigate how the system behaves in the regime where the fields no longer behave like ULDM.

The above estimates show that local constraints on disformal couplings to scalar and vector ULDM imply that disformal couplings could have only played an important role well before BBN. It is remarkable that by using local bounds, one is able to constrain higher derivative interactions in the gravity sector to early times (very high redshifts) and exclude significant disformal couplings during BBN. Note that we have assumed that gravity, in the gravitational metric, is described by the Einstein-Hilbert action. It would be interesting to apply our formalism to general higher derivative theories of gravity. We leave a detailed study on the implications for general modified theories of gravity for future work.

\begin{acknowledgments}
G.D., A.G., and A.T. are supported by the DFG under the Emmy-Noether program, project number 496592360. G.D. also acknowledges support by the JSPS KAKENHI grant No. JP24K00624. N.S.
acknowledges support from the Deutsche Forschungsgemeinschaft under the Heinz
Maier Leibnitz Prize BeyondSM HML-537662082.
F.K. acknowledges support by the German Research Foundation (DFG) under Germany’s Excellence Strategy – EXC-2123 QuantumFrontiers – 390837967.
\end{acknowledgments}

    \appendix

     \section{Transformation using tensor components}
    \label{sec:Transformation_Tensor_Components}

    Here, we derive the transformation of the spin-connection using a component-dependent approach. In the absence of torsion, the spin connection is defined in terms of the tetrads via
    \begin{align}\label{eq:abcappendix}
        \tensor{\omega}{_A_B_C} =  \tensor{e}{_C^\mu} \tensor{e}{_[_A^\nu} \left( \partial_\mu \tensor{e}{_B_]_\nu} - \tensor{e}{_B_]_\lambda} \Gamma^\lambda_{\mu\nu} \right)\,,
    \end{align}
    Employing the transformation rule of the individual tetrads \eqref{eq:ThetaDisformal}, as well as that of the Christoffel symbols, we obtain
    \begin{align}\label{eq:firstomeabc}
        \tensor{\tilde \omega}{_A_B_C} = & - \tensor{\tilde e}{_C^\mu} \tensor{\tilde e}{_[_B^\nu} \left( \partial_\mu \tensor{\tilde e}{_A_]_\nu} - \tensor{\tilde e}{_A_]_\lambda} \tilde \Gamma^\lambda_{\mu\nu} \right)  \nonumber \\
        =& \frac{1}{C} \tensor{({\cal U}^{-1})}{^D_C} \tensor{e}{_D^\mu} \Big[  - \frac{1}{C} \tensor{(\mathcal{U}^{-1})}{^E_[_B} \tensor{e}{_E^\nu} \left( \partial_\mu ( C \tensor{{\cal U}}{_A_]_\nu} )  - C \tensor{{\cal U}}{_A_]_\lambda} \Gamma^\lambda_{\mu\nu} - C \tensor{{\cal U}}{_A_]_\lambda} {\cal K}^\lambda_{\mu\nu} \right) \Big] \nonumber \\
        =& \frac{1}{C} \tensor{({\cal U}^{-1})}{^D_C} \tensor{e}{_D^\mu} \Big[ \tensor{\omega}{_A_B_\mu} - \tensor{(\mathcal{U}^{-1})}{^\nu_[_B} \nabla_\mu \tensor{\mathcal{U}}{_A_]_\nu} + \tensor{(\mathcal{U}^{-1})}{^\nu_[_B} \tensor{\cal U}{_A_]_\lambda} {\cal K}^\lambda_{\mu\nu} \Big]\,.
    \end{align}
    In the first step in Eq.~\eqref{eq:firstomeabc}, we used the definition of the generalized covariant derivative (acting also on local indices), namely
    \begin{align}
        \nabla_\mu \tensor{{\cal U}}{_A_\nu} = \partial_\mu  \tensor{{\cal U}}{_A_\nu} + \tensor{\omega}{_A_B_\mu} \tensor{{\cal U}}{^B_\nu} - \Gamma^{\lambda}_{\mu\nu} \tensor{{\cal U}}{_A_\lambda}~\,.
    \end{align}
    We also introduced for convenience \cite{Zumalacarregui:2013pma}
    \begin{equation}
                \mathcal{K}^\alpha_{\mu\nu}=\tilde \Gamma^\alpha_{\mu\nu} - \Gamma^\alpha_{\mu\nu}=\frac{1}{2}\tilde g^{\alpha\lambda}\left(\nabla_\mu \tilde g_{\nu\lambda}+\nabla_\nu \tilde g_{\mu\lambda}-\nabla_\lambda \tilde g_{\mu\nu}\right)~.
            \end{equation}

    Now, we could either directly use the expression for the transformed metric \eqref{eq:Metric_Disformal_transformation_General} or use the expression of the tilded metric in terms of the tetrads. In the latter, we find that 
    \begin{align}
    \tilde\omega_{ABC}=&\tilde e^\mu_C \tilde  e^\nu_{[A}  \partial_\mu \tilde e_{\nu B]}-\tilde  e^\nu_{C} \tilde e^\mu_{[A}  \partial_\mu \tilde e_{\nu B]}-\tilde e^\mu_{[A} \tilde  e^\nu_{B]}  \partial_\mu \tilde e_{\nu C}\nonumber\\&\qquad\qquad+\Gamma^\alpha_{\mu\nu}\left(\tilde  e^\nu_{C} \tilde e^\mu_{[A} \tilde e_{\alpha B]}+\tilde e^\mu_{[A} \tilde  e^\nu_{B]}   \tilde e_{\alpha C}-\tilde e^\mu_C \tilde  e^\nu_{[A}  \tilde e_{\alpha B]}\right)\,,
    \end{align}
    which is useful to see how the original spin connection is related to the spacetime derivatives of the new untilded metric. In the former, we plug in the expression for the transformed metric \eqref{eq:Metric_Disformal_transformation_General} and find that
    \begin{align}\label{eq:relationskappa}
        \tensor{({\cal U}^{-1})}{^\mu_C}  & \tensor{(\mathcal{U}^{-1})}{^\nu_[_B} \tensor{\cal U}{_A_]_\lambda} {\cal K}^\lambda_{\mu\nu} \nonumber\\&=   2 \eta_{C[A} \tensor{({\cal U}^{-1})}{^\mu_B_]} \nabla_\mu \ln C + \left( \delta^G_C \eta_{E[A} \delta^L_{B]} + \eta_{EC} \delta^L_{[B} \eta_{A]G} \right)  \tensor{({\cal U}^{-1})}{^D_L} \tensor{({\cal U}^{-1})}{^F_G}  \nabla_D \tensor{{\cal U}}{^E_F}\,.
    \end{align}
    By plugging Eq.~\eqref{eq:relationskappa} into Eq.~\eqref{eq:firstomeabc}, we arrive at the final transformation of the spin connection, which is given by
    \begin{align}\label{eq:secondomegaabc}
       \tensor{\tilde \omega}{_A_B_C} = &    \frac{1}{C} \tensor{({\cal U}^{-1})}{^D_C}\, \tensor{\omega}{_A_B_D} + 2 \eta_{C[A} \tensor{({\cal U}^{-1})}{^D_B_]} \nabla_D \ln C \nonumber
                \\ 
                & + \frac{1}{C}\left( \tensor{\delta}{_C^L}\tensor{\delta}{_{[A}^G}\tensor{\delta}{_{B]}^E}
                - \tensor{\delta}{_{[A}^L}\tensor{\delta}{_C^G}\tensor{\delta}{_{B]}^E} - \tensor{\delta}{_{[A}^L}\tensor{\delta}{_{B]}^G}\tensor{\delta}{_C^E} \right)\tensor{(\mathcal{U}^{-1})}{^D_L}\tensor{(\mathcal{U}^{-1})}{^F_G}\,\eta_{EI}\nabla_D\tensor{\mathcal{U}}{^I_F}~.
    \end{align}

    \section{Lorentz transformation}
\label{app:Lorentz_transformation}

By construction, the metric is invariant under Lorentz transformations of the tetrad base
\begin{align}
    \theta^{\prime A} = \tensor{\Lambda}{^A_B} \theta^B
\end{align}
Therefore, the choice of $L$ is not unique. In the following, we will demonstrate how one can use the Lorentz transformation to absorb the antisymmetric components of $L$. Let us split $L = S + A$, where $S$ denotes the symmetric and $A$ the antisymmetric components of $L$.

    \subsection{Perturbative construction of Lorentz transformation}
    Let us parameterize the Lorentz transformation of the tetrad field as
    \begin{align}
        \tensor{\Lambda}{^A_B} = f_0 \delta^A_B + f_1 \tensor{T}{^A_B} + f_2 \tensor{T}{^A_C} \tensor{T}{^C_B} + f_3 \tensor{T}{^A_C} \tensor{T}{^C_D} \tensor{T}{^D_B} \,,
    \end{align}
    where $T$ is an arbitrary anti-symmetric matrix.
    In order to fulfill the Lorentz transformation, we need to require that 
    \begin{align}
        \Lambda^T \eta \Lambda = \eta\,,
    \end{align}
    which leads to the conditions
    \begin{align}
        f_0^2 - f_2^2 \det T - \frac{1}{2} f_3^2 \Trace T^2 \det T = 1, \\
        2 f_0 f_2 - f_1^2 - \frac{1}{2} f_2^2 \Trace T^2 + f_3^2 ( \det T - \frac{1}{4} (\Trace T^2)^2 )  = 0\,,
    \end{align}
    which can be solved for $f_0$ and $f_2$. 
    Defining a new disformal matrix ${\cal U}^\prime$ via
    \begin{align}
        {\cal U}^\prime =  \Lambda {\cal U} = f_0 \mathds{1} + f_0 L + f_1 T + f_1  T L + f_2 T^2 + f_2  T^2 L + f_3 T^3 + f_3 T^3 L\,,
    \end{align}
    we can absorb the linear term in the antisymmetric components $A$ by setting $T_{AB} = A_{AB}$ and $f_1=-f_0$. Further, setting $f_3=-f_2$ removes the cubic term $T^3$.

    The anti-symmetric components of the new disformal transformation are given by
    \begin{align}
       A^\prime  = - f_0 A S - f_2 A^3 S  = {\cal O}(L^2)~.
    \end{align}
    We note that if $L=A$, the new disformal matrix is already symmetric and the construction stops. If $S \neq 0$ we can use the same construction as before and absorb the new anti-symmetric components up to linear order in $L^\prime={\cal O}(L^2)$. Using the procedure iteratively, we can construct a Lorentz transformation that removes all the anti-symmetric components of the disformal transformation. Note that in 4-dimensions the construction stops at the third step by use of the generalized Cayley-Hamilton theorem for the product of two matrices \cite{Generalized_Cayley_Hamilton}.

    \subsection{Redefining spinor field}
Alternatively, the antisymmetric components of $L$ can be absorbed by redefining the spinor field (see also \cite{Kostelecky:2010ze} for a similar discussion). For the sake of simplicity, we only consider the leading-order contributions. Consider the transformation 
            \begin{equation}
                \Psi \to e^{-\frac{i}{2}\tensor{\chi}{_A_B}\tensor{\Sigma}{^A^B}} \Psi
\,.            \end{equation}
        In the following we use $Y= -\frac{i}{2} \chi_{AB} \Sigma^{AB} $ to shorten the notation. 
        Using the identities 
            \begin{align}
                e^{-Y} \gamma^C e^{Y} = & \tensor{\Lambda}{^C_B} \gamma^B \simeq \left( \gamma^C + \tensor{\chi}{^C_F} \gamma^F   \right)\,,  \\
                 \nabla e^Y = & \int_0^1 \md \alpha e^{\alpha Y} \nabla Y e^{(1-\alpha) Y}\,,
            \end{align}
        and substituting it back into the transformed Dirac Lagrangian \eqref{eq:Final_Hermitian_Lagrangian}, we obtain 
        \begin{align}
            S_{\rm Dirac} =  \int \md^4x\, \sqrt{-g}\, & \Big[  \frac{i}{2} \tensor{({\cal U}^{-1})}{^A_B} \tensor{\Lambda}{^B_C} (\bar \Psi \gamma^C \overset{\leftrightarrow}{\nabla}_A \Psi )  +  \frac{i}{2} \tensor{({\cal U})}{^A_B} \bar \Psi  \Big\{ \tensor{\Lambda}{^B_C} \gamma^C, \int \md \alpha e^{-\alpha Y } \nabla_A Y e^{\alpha Y}  \Big\}  \Psi \nonumber \\
    & - M_A \tensor{\Lambda}{^A_B} J_5^B - C \bar \Psi m \Psi \Big]~. \label{eq:Full_Lorentz_transformation}
        \end{align}
        Let us, for simplicity, only focus on the linear order in $L$ and $\chi$ leading to
            \begin{align}
                S_{\rm Dirac} \simeq \int \md^4x\, \sqrt{-g}\, & \Big[ \frac{i}{2} \left( \tensor{\delta}{^A_B} - \tensor{L}{^A_B} + \tensor{\chi}{^A_B} \right)  (\bar \Psi \gamma^B \overset{\leftrightarrow}{\nabla}_A \Psi ) + \frac{1}{4} \nabla_A \chi_{BC} \bar \Psi \{ \gamma^A, \Sigma^{BC} \} \Psi \nonumber \\
                & - \frac{1}{4} \tensor{\epsilon}{_A^B^C^D} \nabla_D L_{BC} J_5^A - C \bar \Psi m \Psi  \Big] \nonumber \\
                \simeq \int \md^4x\, \sqrt{-g}\, & \Big[ \frac{i}{2} \left( \tensor{\delta}{^A_B} - \tensor{L}{^A_B} + \tensor{\chi}{^A_B} \right)  (\bar \Psi \gamma^B \overset{\leftrightarrow}{\nabla}_A \Psi ) - \frac{1}{4} \tensor{\epsilon}{_A^B^C^D} \nabla_D (L_{BC} - \chi_{BC}) J_5^A \nonumber \\
                & - C \bar \Psi m \Psi \Big]~.
            \end{align}
        Therefore, choosing that at leading order that
        \begin{align}
            \tensor{\chi}{^A_B} = \tensor{A}{^A_B} + {\cal O}(L^2)\,,
        \end{align}
        we can absorb the antisymmetric components of the disformal transformation.

 \section{Disformal coupling involving field strength}
    \label{sec:U_!_coupling}

    A straightforward generalization to the disformal transformation includes the field-strength tensor of the vector fields (see, for instance, \cite{DeFelice:2019hxb,Gumrukcuoglu:2019ebp}). If the disformal transformation of the tetrads depends directly on the field strength $F_{\mu\nu}$, e.g., if $L_{\mu\nu}=F_{\mu\nu}$, there seems to be an antisymmetric tensor directly coupled to the kinetic term of the fermions. This is at first a little counterintuitive because the corresponding transformation of the metric does not see the antisymmetric components. As we shall show, such antisymmetric components can be removed via a fermion field redefinition. 
    
     For simplicity, we continue with a single vector field and impose the $U(1)$ gauge symmetry so that the general ansatz for $U$ is given by
    \begin{align}
    \label{eq:Disformal_U1_Naive}
        \tensor{U}{^A_B} = C \tensor{\delta}{^A_B} +D \tensor{\cal F}{^A_B}\,,
    \end{align}
    where $C$ and $D$ are functions of $\Trace {\cal F}^2$ and $\Trace {\cal F}^4$. Note that we could extend the ansatz by including $ {\cal F}^\star$. However, such terms lead to parity violation or are equivalent to the aforementioned terms. Particularly, note that replacing ${\cal F}$ by $ {\cal F}^\star$ leads to the same metric up to a redefinition of the free functions but seemingly to parity violation in the fermionic sector. Next, we may proceed as explained in Sec.~\ref{app:Lorentz_transformation} to absorb the antisymmetric components. 
    
    Another way to convince ourselves that the antisymmetric components of the tetrad disformal transformation given by Eq.~\eqref{eq:Disformal_U1_Naive} are not physical is by noting that there is another tetrad disformal transformation which leads to the same metric disformal transformation. By using the Cayley Hamilton theorem, one can show that the tetrad disformal transformation given by
    \begin{align}
        \tensor{{U^\prime}}{^A_B} = C^\prime \tensor{\delta}{^A_B} +  D^\prime \tensor{{\cal F}}{^A_C} \tensor{{\cal F}}{^C_B}\,,
    \end{align}
    leads to the same metric as Eq.~\eqref{eq:Disformal_U1_Naive} when
    \begin{align}
        C^2 = C^{\prime 2} (1 - D^{\prime 2}\, {\rm det}{\cal F}), \qquad D^2 = - D^\prime ( 2 - \frac{1}{2} D^\prime \Trace {\cal F}^2 ) \,.
    \end{align}
   The reason why the two tetrad transformations lead to the same metric transformation is that they are related via a Lorentz transformation of the tetrad field, or alternatively, after a field redefinition. It is interesting to note that $C^\prime$ is a non-trivial function depending on both $C$ and $D$, and, therefore, the relation between both choices mixes the conformal and disformal transformation. 

    Lastly, we compute the leading order in ${\cal F}_{\mu\nu}$ in the transformed Dirac action. By using both  ${\cal U}$ (after field redefinition) or ${\cal U}^\prime$, we find that 
    \begin{align}
        {\cal L}^\prime \simeq {\cal L}_{\rm Dirac} - \frac{i}{2} \tensor{\cal F}{^A_C} \tensor{{\cal F}}{^C_B} (\bar \Psi \gamma^B \overset{\leftrightarrow}{\nabla}_A \Psi) - m  C^\prime_{,\Trace {\cal F}^2} \Trace {\cal F}^2   \bar \Psi \Psi + {\cal O}(F^3)\,,
    \end{align}
    where we also expanded the conformal factor at leading order in $\Trace {\cal F}^2$. Note that the coupling to the chiral current is of order $M_A = { \cal O}(L^{\prime 2}) = {\cal O}({\cal F}^4)$. Starting from ${\cal U}$, it is highly non-trivial to see that up to cubic order ${\cal O}({\cal F}^3)$ all terms can be canceled via a field redefinition. 
    Note that the leading-order correction to the Dirac equation is of order ${\cal O}({\cal F}^2)$ and, therefore, of the same order as for the bosons or gravity sector, in contrast to the naive expectation starting from Eq. \eqref{eq:Disformal_U1_Naive}. Nevertheless, in a separate publication \cite{auth:us2}, we show that the disformal transformation studied in this subsection actually leads to higher derivatives and possible ghosts due to the axial coupling.

\section{Disformal coupling to vector bosons \label{app:bosons}}
For completeness, let us shortly discuss the disformal coupling to $U(1)$ gauge bosons and the transformation of the action. For related literature see Refs.~\cite{Domenech:2018vqj,Jirousek:2018ago,DeFelice:2019hxb,Minamitsuji:2020jvf,Hell:2024xbv}. Consider a $U(1)$ gauge field whose action reads 
    \begin{align}
        {\cal L}_{\rm EM} = & - \frac{1}{4} \int\md x^4 \sqrt{-\breve g}\,\breve g^{\mu\alpha} \breve g^{\nu\beta} F_{\alpha\beta}  F_{\mu\nu}  \,.
    \end{align}
    Using that $\breve g^{\mu\nu}=\tensor{({\cal U}^{-1})}{_\alpha^\mu}\tensor{({\cal U}^{-1})}{_\beta^\nu}g^{\alpha\beta}$ (see Eqs.~\eqref{eq:ThetaDisformalInv} and \eqref{eq:Metric_Disformal_transformation_General}), and obtain that the gauge field action after the transformation is given by
    \begin{align}
    \label{disformalgaugeL}
        S_{\rm EM} 
        = & - \frac{1}{4} \int \md^4x\,\sqrt{-g} \,G^{\alpha\mu\beta\nu}F_{\alpha\beta} F_{\mu\nu}\,,
    \end{align}
    where we introduced an area metric $G^{\alpha\mu\beta\nu}$ given by
    \begin{align}
    \label{disformalgaugeG}
    G^{\alpha\mu\beta\nu}=\det  \mathcal{U} \,
    \tensor{({\cal U}^{-1})}{^\alpha_\sigma} \tensor{({\cal U}^{-1})}{^\sigma^\mu} \tensor{({\cal U}^{-1})}{^\beta_\lambda} \tensor{({\cal U}^{-1})}{^\lambda^\nu}\,.
    \end{align}
    Note that the conformal factor does not appear due to the well-known conformal invariance of the gauge field action in four dimensions. The disformal part shifts the kinetic term of the photons and, in particular, it modifies the propagation speed. Please also note that while we abused the notation ${\cal U}^{-1}$ for the disformal transformation, it is not necessarily the same tensor as for $\tilde g$. We show one example below.
    
    Since we mainly focus on the scalar and vector disformal couplings, we explicitly write the transformation of the gauge field action for this case. Let us assume that the disformal transformation is given by
\begin{align}\label{eq:disformalappendix}
    \breve g_{\mu\nu}=&\Omega^2\left(g_{\mu\nu}+{\cal K}\, B_\mu B_\nu\right)=\Omega^2(\tensor{\delta}{^\alpha_\mu}+K B^{\alpha} B_\mu)(\tensor{\delta}{^\beta_\nu}+K B^{\beta} B_\nu)g_{\alpha\beta}\,,
    \end{align}
    where ${\cal K}=K(2+KX)$. In that case, we have that 
    \begin{align}
    \tensor{({\cal U}^{-1})}{^\mu_\nu} = \delta^\mu_\nu-\frac{K}{1+ K X} B^{\mu} B_\nu\,,
    \end{align}
    and, therefore, it follows that
    \begin{align}
    \label{disformalgaugeG2}
    G^{\alpha\mu\beta\nu}&=\left(1+KX\right)\left(g^{\alpha\mu}g^{\beta\nu}-\frac{2K(2+KX)}{(1+KX)^2}\left(g^{\alpha\mu}B^{\beta}B^{\nu}+B^{\alpha}B^{\mu}g^{\beta\nu}\right)\right)\nonumber\\&=\sqrt{1+{\cal K} X} \left(g^{\alpha\mu}g^{\beta\nu}-\frac{2{\cal K}}{1+{\cal K} X}\left(g^{\alpha\mu}B^{\beta}B^{\nu}+B^{\alpha}B^{\mu}g^{\beta\nu}\right)\right)\,,
    \end{align}
    where we dropped a term proportional to $B^\alpha B^\mu B^\beta B^\nu$ because of symmetries. Namely, it vanishes after contraction with the field strength.

    In the main text, we used GW observations \cite{LIGOScientific:2017vwq,LIGOScientific:2017zic} to fix $\breve g_{\mu\nu}=g_{\mu\nu}$ so that photons and GWs propagate at the same speed. We have, however, a relative modification between the electromagnetic field and fermions. We look at this in more detail in App.~\ref{app:bosons}.

\section{Constraints from isotropic Lorentz violations \label{app:isolorentzviolations}}

The constraints given in Eq.~\eqref{neutronlimits} do not apply to the case of isotropic Lorentz violation. To test for isotropic Lorentz violation, we  make use of SME bounds by the BASE (Baryon Antibaryon Symmetry Experiment) collaboration, which runs a Penning-trap experiment at
CERN's Antiproton Decelerator facility has measured the charge-to-mass ratios of the proton and the antiproton with a fractional uncertainty of 16 parts per trillion. The sampling period in the BASE experiment was 4 min, and data were taken for 1.5 years~\cite{Wursten:2023bsl}.
In the presence of SME coefficients, the energy levels of the trapped particles are expected to be shifted, leading to modified cyclotron frequencies. Applying the formalism developed in Ref.~\cite{Ding:2020aew}, they obtain the following bounds on the spin-independent coefficients~\cite{BASE:2022yvh,Wursten:2023bsl}:
\begin{align}
\begin{split}
    |\tilde{c}_e^{XX}|\,, |\tilde{c}_e^{YY}| &< 7.79 \times 10^{-15}\,,\\
    |\tilde{c}_e^{ZZ}| &< 4.96 \times 10^{-15}\,,\\
    |\tilde{c}_p^{XX}|\,, |\tilde{c}_p^{*XX}|\,,
    |\tilde{c}_p^{YY}|\,, |\tilde{c}_p^{*YY}| &<2.86 \times 10^{-11}\,,\\
    |\tilde{c}_p^{ZZ}|\,, |\tilde{c}_p^{*ZZ}| & <1.82 \times 10^{-11}    \,.
\end{split}
\label{eq:BASElims}
\end{align}
Taking only the $c_{\mu\nu}$ coefficients to be nonzero, these translate into constraints on $\tilde{c}_f^{JJ}\to c_f^{JJ}$, with flavors $f=e,p$ and $JJ=XX, YY, ZZ$, and similar for the starred quantities, which were obtained for the antiproton~\cite{Ding:2020aew}. Since the $c_{\mu\nu}$ coefficients are CPT-even, these are expected to be equal.
More stringent constraints can be obtained from limits on modified energy thresholds, which would lead to effects such as vacuum \v{C}erenkov radiation or photon decay~\cite{PhysRevD.1.961,Altschul:2006zz,Coleman:1997xq,Lehnert:2004be,Hohensee:2008xz,Klinkhamer:2008ky}, see e.g. Ref.~\cite{Crivellin:2022idw}.\footnote{Assuming isotropic Lorentz violation, i.e. $c_{\mu\nu}$ coefficients of the form $c^{\mu\nu} = \mathring{c}~\mathrm{diag}\left(1,\frac{1}{3},\frac{1}{3},\frac{1}{3}\right)$ the following bounds are derived for electrons and muons:
$-1.6\times 10^{-14} < \mathring{c}^{(\mathrm{e})} < 1.9\times 10^{-19}$, $-4.2 \times 10^{-15} < \mathring{c}^{(\mu)} < 8.3\times 10^{-15}$.
Taking the best bounds across these two flavours, this leads to
$-4.2 \times 10^{-15} < \mathring{c} < 1.9\times 10^{-19}$.
} However, since these bounds are derived from single events, they cannot be used to constrain a slowly varying signal.

Let us now apply these bounds to the case of unpolarized vector ULDM (see discussions around Sec.~\ref{ssec:constraints} for more details). In that case, the vector field is statistically isotropic in the range of the experiment, namely  $\langle B_iB_j\rangle \sim \frac{1}{3}\delta_{ij}$, and we have that Eq.~\eqref{eq:c_vec_general} reads
\begin{align}
\label{eq:isotropiccmunu}
    \big[\langle c_{\mu\nu}\rangle_{T}\big] = \frac{\rho_{\rm DM}D}{m_B^2-\frac{3}{4} \rho_{\rm DM}D} \left[\frac{1}{4} 
    \mathrm{diag}\left( 1, \frac{1}{3}, \frac{1}{3}, \frac{1}{3}\right)
    \right]_{\mu\nu}\,,
\end{align}
where we worked at leading order in $|\vec{v}|$. Using the most stringent bounds from experimental results~\cite{BASE:2022yvh,Wursten:2023bsl} listed in Eq.~\eqref{eq:BASElims}, we obtain
\begin{align}
\label{isovectDLambdaconstraints}
\begin{split}
    &D \lesssim 3\times 10^{-20}\;\text{GeV}^{-2}\left(\frac{m_B}{10^{-15}\,{\rm eV}}\right)^{2}\\
&\Lambda \gtrsim  6 \times 10^{9}\;\text{GeV}\left(\frac{m_B}{10^{-15}\,{\rm eV}}\right)^{-1}
\end{split}\quad \text{(unpolarized vector ULDM)}\,.
\end{align}
This constraint on the new physics scale $\Lambda$ is approximately seven orders of magnitude weaker than that in Eq.~\eqref{vectorDLambdaconstraints}, but still fourteen orders of magnitude stronger than the constraint on $\Lambda$ in the scalar case (Eq.~\eqref{scalarDLambdaconstraints}). 
The high sensitivity to the disformal coupling $D$ in the case of vector ULDM results from the combination of a lack of velocity suppression, a sensitivity scaling with $1/m_B^2$, and a dimension-six operator rather than a dimension-eight operator. 
The contrast between the bounds in Eq.~\eqref{vectorDLambdaconstraints} and Eq.~\eqref{isovectDLambdaconstraints} strengthens the case for a better modelisation of the dark photon polarisation distribution.

\section{Constraints from boost violations \label{app:boostviolations}}

Throughout this work, we have assumed that photons and gravitational waves couple to the same metric to satisfy constraints on their relative propagation speeds from GW observations \cite{LIGOScientific:2017vwq,LIGOScientific:2017zic}. However, although GWs and photons propagate at the same speed, there is a relative difference between the electromagnetic field and fermions, since we assumed that they couple to different metrics.

As an alternative formulation, we could leave fermions minimally coupled to $\tilde g_{\mu\nu}$ and instead do the inverse disformal transformation to the gauge field action, as given in App.~\ref{app:bosons}. The corresponding transformation would be given by setting $\breve g_{\mu\nu}\to g_{\mu\nu}$ and $g_{\mu\nu}\to \tilde g_{\mu\nu}$, with $K=-D$, in Eq.~\eqref{eq:disformalappendix} and Eq.~\eqref{disformalgaugeG2}. With that, we obtain
 \begin{align}
    \label{disformalgaugeL2}
        S_{\rm EM} 
        = & - \frac{1}{4} \int \md^4x\,\sqrt{-\tilde g} \,\tilde G^{\alpha\mu\beta\nu}F_{\alpha\beta} F_{\mu\nu}\,,
    \end{align}
where 
  \begin{align}
    \label{disformalgaugeG3}
    \tilde G^{\alpha\mu\beta\nu}&=\left(1-D\tilde X\right)\left(\tilde g^{\alpha\mu}\tilde  g^{\beta\nu}+\frac{2D(2-D\tilde X)}{(1-D\tilde X)^2}\left(\tilde  g^{\alpha\mu}B^{\beta}B^{\nu}+B^{\alpha}B^{\mu}\tilde  g^{\beta\nu}\right)\right)\,,
    \end{align}
    where $\tilde X=\tilde g_{\mu\nu}B^\mu B^\nu$. With Eq.~\eqref{disformalgaugeG3}, we can discuss bounds on the disformal coupling to gauge bosons. For related literature see Refs.~\cite{Domenech:2018vqj,Jirousek:2018ago,DeFelice:2019hxb,Minamitsuji:2020jvf,Hell:2024xbv}. 
    
    Before that, it is interesting to note that the disformal transformation discussed above resulting in Eq.~\eqref{disformalgaugeL2}, can also be understood, in flat spacetime (that is $\tilde g_{\mu\nu}=\eta_{\mu\nu}$), as a coordinate transformation given by $x^\mu \rightarrow x^\mu - \tensor{c}{^\mu_\nu}x^\nu$. This leaves the physics invariant, yet naively appears to completely remove $c_{\mu\nu}$ from Eq.~\eqref{LSME}, leaving a conventional Dirac theory with metric $\eta_{\mu\nu}$. However, the covariant derivative $D_\mu$ includes the photon gauge field $A^\mu$, and the full gauge theory includes the kinetic term 
$-\tfrac{1}{4}\eta_{\mu\nu}\eta_{\rho\sigma}F^{\mu\rho}F^{\nu\sigma}$ with (by initial assumption) metric $\eta_{\mu\nu}$. Since we used bounds derived within the SME formalism, we shall use their notation in what follows, having in mind that it is equivalent to the inverse disformal transformation.

In this way, one finds that the coordinate transformation actually shifts the Lorentz violation into the photon-sector metric $\eta_{\mu\nu} \rightarrow \breve{g}_{\mu\nu} = \eta_{\mu\nu} - 2c_{\mu\nu}$~\cite{Kostelecky:2002hh,Bailey:2004na, Altschul:2006zz}. 
A metric of this type is equivalent to a subset of the photon-sector term
\begin{equation}
\label{photonLV}
{\cal L}_{\rm SME}^{\rm photon} \supset -\frac{1}{4}k_{F}^{\mu\nu\rho\sigma}F_{\mu\nu}F_{\rho\sigma},
\end{equation}
where $k_{F}^{\mu\nu\rho\sigma}$ has the symmetries of the Riemann tensor, is double traceless, and has 19 independent components. Note the correspondence with the analogous disformal gauge terms in Eqs.~\eqref{disformalgaugeL2} and \eqref{disformalgaugeG3}.
The nine "Ricci-like" components may be decomposed as~\cite{Altschul:2019beo}
\begin{equation}
k_F^{\mu\nu\rho\sigma} = \frac{1}{2}\left(\eta^{\mu\rho}\widetilde{k}_F^{\nu\sigma} - \eta^{\mu\sigma}\widetilde{k}_F^{\nu\rho} - \eta^{\nu\rho}\widetilde{k}_F^{\mu\sigma} + \eta^{\nu\sigma}\widetilde{k}_F^{\mu\rho}\right),
\end{equation}
where $\widetilde{k}_F^{\mu\nu} \equiv \tensor{k}{_F_\alpha^\mu^\alpha^\nu}$ and  $\eta_{\mu\nu}\widetilde{k}_F^{\mu\nu} = 0$. The correspondence with Eq.~\eqref{disformalgaugeG3} is given by identifying 
\begin{align}
\widetilde{k}_F^{\mu\nu}\leftrightarrow \frac{2D(2-DX)}{1-DX}\left(B^\mu B^\nu-\frac{X}{4} \eta^{\mu\nu}\right)\,,
\end{align}
where we already took $\tilde  g^{\mu\nu}= \eta^{\mu\nu}$ when matching.
Note, however, that in the inverse disformal transformation the standard kinetic term $F_{\mu\nu}F^{\mu\nu}$ is still multiplied by a factor $1-DX$, which could be absorbed in a redefinition of the charge.

Assuming the remaining 10 "Weyl-like" components are set to zero, one may thus view the photon metric as perturbed from Minkowski by $\widetilde{k}_F^{\mu\nu}$, or $\widetilde{k}_F^{\mu\nu}-2c^{\mu\nu}$ if Lorentz violation is initially introduced in the fermion sector. Conversely, a coordinate transformation $x^\mu \rightarrow x^{\mu} - \tfrac{1}{2}(\widetilde{k}_{F})^{\mu}_{\phantom{\mu}\nu} x^\nu$ can be performed to render the photon metric Minkowski and fermion metric $\eta
^{\mu\nu}  + c^{\mu\nu}-\tfrac{1}{2}\widetilde{k}_F^{\mu\nu}$.\footnote{Note this implies the constraints~\eqref{neutronlimits} from Ref.~\cite{Smiciklas:2011xq} have implicitly set the photon coefficients to zero.} Regardless of the (unphysical) choice of coordinates,  observables only depend on the relative difference between the fermion and photon metrics.

Among the physical effects implied by $\widetilde{k}_F^{\mu\nu}$ are orientation-dependent modifications of the photon phase and group speeds  $v_{\rm p} = v_{\rm g} = 1-\tfrac{1}{2}\widetilde{k}_F^{\mu\nu}\hat{p}_\mu \hat p_\nu$, where $\hat{p}^\mu \equiv p^\mu/|\vec{p}|$. Constraints up to the $\sim 10^{-20}$ level have been extracted on the rotation-invariant (isotropic) combination of components $\widetilde{\kappa}_{\rm tr}$ and $\left(\widetilde{\kappa}_{\rm tr}-\tfrac{4}{3}c^e_{00}\right)$~\cite{auth:KosteleckTables2025}. These are thus constraints on violations of Lorentz boost invariance. Analogous effects also result from disformal interactions~\cite{vandeBruck:2016cnh}. 

Let us investigate the implications for the disformal coupling to fermions studied in the main part of this work. For simplicity, we focus on the conservative case of isotropic Lorentz violation (see discussion around Eq.~\eqref{eq:isotropiccmunu}), where the SME coefficient $\widetilde{k}_F^{\mu\nu}$ is written as $\widetilde{k}_F^{\mu\nu} = \frac{3}{2}\tilde{\kappa}_{\rm tr} \mathrm{diag}\left(1, \frac{1}{3},\frac{1}{3},\frac{1}{3}\right)$ with $\tilde{\kappa}_{\rm tr}$ parametrizing the strength of the Lorentz-violating effects. In particular, the modification to the photon speed is given by $v_p \approx 1-\widetilde{\kappa}_{\rm tr}$~\cite{Duenkel:2021gkq,Klinkhamer:2008ky}. In what follows, we will work to first order in the disformal coupling and focus on matching the average value of $\widetilde{k}_F^{\mu\nu}$, and so $\tilde{\kappa}_{\rm tr}$, to $D$. 

For the scalar field, we find that 
\begin{align}
\langle\widetilde{k}_F^{\mu\nu} \rangle=\frac{2D\langle X\rangle(2-D\langle X\rangle)}{1-D\langle X\rangle}\left(\langle v^\mu v^\nu\rangle-\frac{1}{4} \eta^{\mu\nu}\right)\,,
\end{align}
where we introduced $v_\mu=\partial_\mu\phi/\sqrt{X}$. Since we have that $\langle X\rangle\approx \rho_{\rm DM}$ and $v_\mu\approx (1,\vec{v})$, we obtain $D^S=\tilde{\kappa}_{\rm tr}/(2\rho_{\rm DM})$, at leading order in $|\vec{v}|\ll 1$. For the vector case, we proceed similarly, namely, we compute
\begin{align}
\langle\widetilde{k}_F^{\mu\nu} \rangle=\frac{2D\langle X\rangle(2-D\langle X\rangle)}{1-D\langle X\rangle}\left(\langle \epsilon^\mu \epsilon^\nu\rangle+\frac{1}{4} \eta^{\mu\nu}\right)\,,
\end{align}
and use that $\langle X\rangle=-\rho_{\rm DM}/m_B^2$ and $\langle \epsilon_i \epsilon_j\rangle\sim \delta_{ij}/3$, where $\epsilon_0\ll |\epsilon_i|$. From this, we conclude that $D^V={3m_B^2}\tilde{\kappa}_{\rm tr}/(2\rho_{\rm DM})$. We may now place bounds on $D$ for the scalar and vector ULDM cases.

The SME data tables list multiple bounds on $\tilde{\kappa}_{\rm tr}$ from tabletop experiments, which involve a continuous data-taking process and can thus be used to constrain the amplitude of a time-varying field. 
A number of these bounds are at the order of $10^{-9}$~\cite{auth:KosteleckTables2025}. 
For the sake of clarity, we use a bound obtained by a Michelson-Morley-type experiment that was rotated at a period of 100~s to correct for potential asymmetries in the experimental setup: The $1~\sigma$ confidence interval yields $ |\tilde{\kappa}_{\rm tr}| \lesssim  10^{-9}$~\cite{Nagel:2014aga}. This corresponds, for scalar ULDM, to a sensitivity 
\begin{align}
\begin{split}
    &D\lesssim 10^{32}\,{\rm GeV}^{-4}\\
&\Lambda \gtrsim 8\,{\rm eV}
\end{split}\quad \text{(scalar ULDM)}\,,  
\end{align}
which is approximately 3 orders of magnitude weaker than the bound derived in Eq.~\eqref{scalarDLambdaconstraints}. For vector ULDM, we obtain
\begin{align}
\begin{split}
    &D \lesssim 10^{-15}\;\text{GeV}^{-2}\left(\frac{m_B}{10^{-15}\,{\rm eV}}\right)^{2}\\
&\Lambda \gtrsim  10^{7}\;\text{GeV}\left(\frac{m_B}{10^{-15}\,{\rm eV}}\right)^{-1}
\end{split}\quad \text{(unpolarized vector ULDM)}\,,
\end{align}
about three orders of magnitude weaker than Eq.~\eqref{isovectDLambdaconstraints} and ten orders of magnitude weaker than Eq.~\eqref{vectorDLambdaconstraints}. Although weaker bounds, they are nevertheless complementary to those discussed in the main text.

Note that bounds on vacuum \v{C}erenkov radiation and photon decay can provide much stronger laboratory bounds on $\tilde{\kappa}_{\rm tr}$ and the equivalent coefficient in the fermion sector~\cite{Altschul:2009xh,Diaz:2013wia,Stecker:2013jfa,Stecker:2014xja}. These are based on single detection events and cannot be used to constrain a time-varying background field. They can, however, be used to constrain an effectively static field, i.e., one that only varies on cosmological time scales like dark energy. This scenario was discussed, e.g., in Ref.~\cite{vandeBruck:2016cnh} in the context of a quintessence dark energy model (note $(\rho_{\rm DM}/\rho_{\rm DE})^{1/4}\approx 30$).
Using the most stringent bounds derived in Ref.~\cite{Crivellin:2022idw}, we obtain sensitivities to $\Lambda$ at the keV scale for disformal couplings to dark energy.

    \bibliography{refs.bib}

 \end{document}